\documentclass[aps,pra,twocolumn,floatfix,showpacs,preprintnumbers,amsmath,amssymb,10pt]{revtex4-1}

\usepackage[dvips]{graphics}
\usepackage{enumitem}

\DeclareMathAlphabet{\mathsfit}{\encodingdefault}{\sfdefault}{m}{sl}
\SetMathAlphabet{\mathsfit}{bold}{\encodingdefault}{\sfdefault}{bx}{sl}

\usepackage[colorlinks,pdfusetitle,urlcolor=blue,citecolor=blue,linkcolor=blue,bookmarksnumbered,plainpages=false]{hyperref}
\usepackage[dvips]{graphics}
\usepackage{geometry}
\DeclareFontEncoding{LS1}{}{}
\DeclareFontSubstitution{LS1}{stix}{m}{n}
\DeclareSymbolFont{symbols2}{LS1}{stixfrak}{m}{n}
\DeclareMathSymbol{\typecolon}{\mathbin}{symbols2}{"25}
 \usepackage[percent]{overpic}

\usepackage{mathptmx}
\usepackage{psfrag}
\usepackage{graphicx}
\usepackage{bm}
\usepackage{amsmath}
\usepackage{trfsigns}
\usepackage{amssymb}
\usepackage{subfigure}
\usepackage[usenames,dvipsnames]{color}
\usepackage{dashrule}
\usepackage{textgreek}
\usepackage[english]{babel}
\usepackage{contour}
\usepackage{placeins}
\usepackage{upgreek,bm}
\usepackage{booktabs}
\usepackage{soul}
\usepackage{color}
\definecolor{dred}{rgb}{.6,.0,0.}
\definecolor{dblue}{rgb}{.0,.0,0.6}
\usepackage{mathtools}

\renewcommand{\vec}[1]{\mathbf{#1}}

\usepackage{xcolor}
\newcommand{\tens}[1]{\mbox{\textsf{\textbf{#1}}}}

\newcommand{\Greektens}[1]{{\bm{#1}}}

\usepackage{lmodern}

\newcommand{\dif}{\mathrm{d}}
\newcommand{\mi}{\textrm{i}} 
\newcommand{\me}{\mathrm{e}}
\newcommand{\B}{\mathbf}
\newcommand{\rSigma}{R_\sigma}

\newcommand{\kPerp}{k_\perp}
\newcommand{\rPerp}{r_\perp}

\newcommand{\beamWaist}{\textrm{w}}
\newcommand{\beamwaist}{\textrm{w}}

\begin{document}

\title{Probing the Purcell effect without radiative decay: Lessons in the frequency and time domains }

\author{Frieder Lindel$^1$}
\author{Francesca Fabiana Settembrini$^2$}
\author{Robert Bennett$^{3}$}
\author{Stefan Yoshi Buhmann$^{4}$}
\affiliation{$^1$ Physikalisches Institut, Albert-Ludwigs-Universit\"at Freiburg, Hermann-Herder-Stra{\ss}e 3, 79104 Freiburg, Germany\\
$^2$ ETH Zurich, Institute of Quantum Electronics, Zurich, Switzerland \\
$^3$ School of Physics and Astronomy, University of Glasgow, Glasgow G12 8QQ, United Kingdom \\
$^4$ Institut f\"ur Physik, Universit\"at Kassel, Heinrich-Plett-Stra{\ss}e 40, 34132 Kassel, Germany
}

\date{\today}

%

\begin{abstract}
The microscopic processes underlying electro-optic sampling of quantum-vacuum fluctuations are discussed, leading to the interpretation of these experiments in terms of an exchange of virtual photons. With this in mind it is shown how one can directly study the Purcell effect, i.e. the changes induced by cavities upon the quantum vacuum, in the frequency and time domains. This forges a link between electro-optic sampling of the quantum vacuum and geometry-induced vacuum effects.
\end{abstract}

\maketitle

\begin{figure}[b]
\centering
\includegraphics[width =\columnwidth]{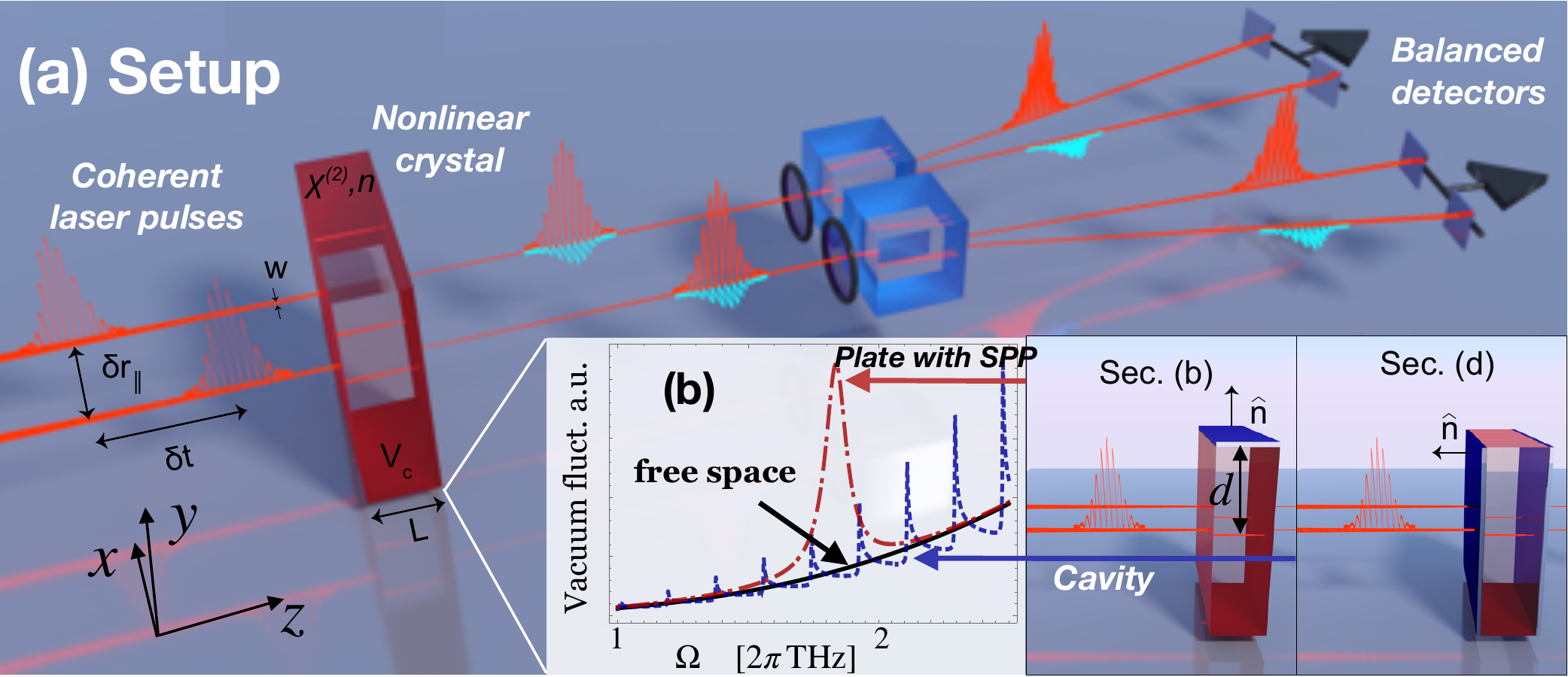} 
\caption{(a) \textit{Electro-optic sampling.} Basic setup: Two laser pulses with tuneable spatial offset $\delta \vec{r}_\parallel$ and delay $\delta t$ mix with vacuum fluctuations inside a nonlinear crystal through its nonlinear susceptibility $\chi^{(2)}$ such that the outgoing pulses contain information about the quantum vacuum inside the crystal. (b) \textit{Purcell effect.} Spectrum of the coincidence limit of the vacuum-correlation function in the bulk of a ZnTe-crystal with infinite extension (solid line), at a distance $d= 17\,\mu$m to a plate with a surface-plasmon polariton at $\Omega = 1.85\,2\pi$THz (dotted-dashed) and inside a cavity of length $L = 0.5\,$mm (dashed). }
\label{fig:SetupWhat}
\end{figure}

Ground-state fluctuations of the electromagnetic field can be seen as responsible for observable effects such as the Lamb shift \cite{Lamb1947}, Casimir forces \cite{Casimir1948} or spontaneous emission \cite{Milonni1994}. Just as the ground-state fluctuations can alter states of matter, matter can in turn influence the quantum vacuum: ground-state fluctuations in close proximity to macroscopic objects---the so-called polaritonic or medium-assisted quantum vacuum---are inherently different from their free-space counterpart. This is the essence of the Purcell effect, see Fig.~\ref{fig:SetupWhat}~(b). Shaping the quantum vacuum using optical environments, such as cavities or surface plasmon-polaritons, can thus be exploited to e.g. enhance or suppress spontaneous emission rates \cite{Purcell1946} or resonant energy transfer \cite{Hemmerich2018}. In the strong coupling regime it even allows the alteration of chemical properties of molecules, for example their reaction rates \cite{hutchison2012modifying,ribeiro2018polariton}.

A novel route to studying ground-state fluctuations of the electromagnetic field has been introduced by means of electro-optic sampling (EOS) experiments \cite{Riek2015,Benea-Chelmus2019,Seletzkiy}, see Fig.~\ref{fig:SetupWhat}~(a). In the setup described in Ref.~\cite{Benea-Chelmus2019} two linearly-polarised, ultra-short laser pulses propagate through a nonlinear crystal, in which they couple to vacuum fluctuations of the electromagnetic field via the nonlinear susceptibility inside the crystal. This coupling leads to a change of the polarisation direction of the laser pulses, which can be measured to infer information about the quantum vacuum inside the crystal on a sub-cycle time scale. This has been used to measure bare \cite{Riek2015,Moskalenko2015} as well as squeezed \cite{riek2017subcycle,kizmann2019subcycle} quantum-vacuum noise. By tuning the temporal and spatial shifts between the two laser pulses, one can additionally detect vacuum correlations between distinct spatio-temporal regions and this way access the spectrum of the electromagnetic ground state \cite{Benea-Chelmus2019}, making EOS a promising tool for an in-depth study of medium-assisted vacuum fluctuations \cite{lindel2020theory,PRAlong}. 

In this work it is shown how EOS can be used to directly access environment-induced changes of the quantum vacuum. By interpreting ground-state correlations as the exchange of virtual photons, it is shown how one can observe the polaritonic quantum vacuum in the time domain. This allows one to study the dynamical formation of cavity modes in the quantum vacuum via multiple reflections, leading to a time-frequency uncertainty relation between the frequency resolution of the observed vacuum correlations and the time during which the quantum vacuum is observed. \\ 
To do so, we build upon previous theoretical results which have been introduced and compared  in Ref.~\cite{lindel2020theory,PRAlong} to experimental data without consideration of the Purcell effect. The EOS signal accounting for absorption and dispersion inside the nonlinear crystal, as well as allowing for general optical environments and pump-pulse profiles, was found to be given by:
\begin{multline} \label{eq:S2ofFilter}
g(\delta t, \delta \vec{r}_\parallel)  = \int_{V_\textrm{C}} \!\! \dif^3 r \!\! \int_{V_\textrm{C}} \!\! \dif^3 r^\prime\!\! \int_0^\infty \hspace{-0.3cm} \dif \Omega \int_0^{\infty} \hspace{-0.3cm}\dif \Omega^\prime \\
\times F(\vec{r},\vec{r}^\prime, \Omega, \Omega^\prime,\delta \vec{r}_\parallel, \delta t) \langle \hat{ E}_{\mathrm{vac},x}(\vec{r},\Omega) \hat{E}_{\mathrm{vac},x}^\dagger(\vec{r}^\prime,\Omega^\prime) \rangle. 
\end{multline}
Here, $ F(\vec{r},\vec{r}^\prime, \Omega, \Omega^\prime,\delta \vec{r}_\parallel, \delta t)$ is a filter function depending on the spatio-temporal shape of the two laser pulses as well as on their lateral and temporal shifts $\delta \vec{r}_\parallel$ and $\delta t$, respectively, see Ref.~\cite{SUPP}, and $V_\textrm{C}$ is the volume of the crystal.
$\langle \hat{ \vec{E}}_\mathrm{vac}(\vec{r},\Omega) \hat{\vec{E}}_\mathrm{vac}^\dagger(\vec{r}^\prime,\Omega^\prime) \rangle$ is the two-point ground-state correlation function of the electromagnetic field in an absorptive and dispersive optical environment, provided by macroscopic QED as \mbox{$\langle \hat{ \vec{E}}_\mathrm{vac}(\vec{r},\Omega) \hat{\vec{E}}_\mathrm{vac}^\dagger(\vec{r}^\prime,\Omega^\prime) \rangle = \frac{\hbar \Omega^2 }{c^2 \varepsilon_0\pi} \delta (\Omega-\Omega^\prime) \mathrm{Im} \tens{G}(\vec{r},\vec{r}^\prime,\Omega)$}  \cite{Buhmann2012BothBooks}. 
Here, $\varepsilon_0$ is the free-space permittivity, $c$ the speed of light in free space and $\tens{G}(\vec{r},\vec{r}^\prime, \Omega)$ is the dyadic Green's function describing the propagation of a photon at frequency $\Omega$ from $\vec{r}^\prime$ to $\vec{r}$ \cite{SUPP}. $\tens{G}$ and thus $\langle \hat{ \vec{E}}_\mathrm{vac}^2 \rangle$ depend on the geometry of the optical environment [Fig.~\ref{fig:SetupWhat}~(b)]. Equation \eqref{eq:S2ofFilter} suggests that by tuning the laser pulses and thus $F$, one can use EOS to access various characteristics of the two-point correlation functions of the electromagnetic ground state.

Throughout this paper we assume that all involved laser pulses are linearly polarised and Gaussian, with beam waist $\beamwaist = 80\,\mu$m, 
 central frequency $\omega_\mathrm{c}= 375 \times 2\pi \,\mathrm{THz}$ and pulse duration $\Delta t = 80\,\mathrm{fs}$ \cite{SUPP}. Unless stated otherwise, we consider the nonlinear crystal of length $L =0.1 \,\mathrm{mm}$ to be a ZnTe crystal whose optical characteristics are summarized in the supplementary material \cite{SUPP}.

\paragraph{Theory}

In order to show how medium-induced changes of the quantum vacuum can be found in EOS experiments, we consider a plate and a cavity with different orientations, attached to the nonlinear crystal (see bottom right of Fig.~\ref{fig:SetupWhat}a). In the presence of additional surfaces the Green's tensor splits into its bulk part $\tens{G}^{(0)}$ and scattering part $\tens{G}^{(1)}$ with $\tens{G}=\tens{G}^{(0)}+ \tens{G}^{(1)}$. The term $\tens{G}^{(1)}$ accounts for all reflection effects, such that restricting $\tens{G}$ to $\tens{G}^{(0)}$ is equivalent to neglecting all influences of any macroscopic object near to the nonlinear crystal as well as reflections at the surfaces of the crystal itself. 
%
\begin{figure}
  \centering
   \includegraphics[width = 1.\columnwidth]{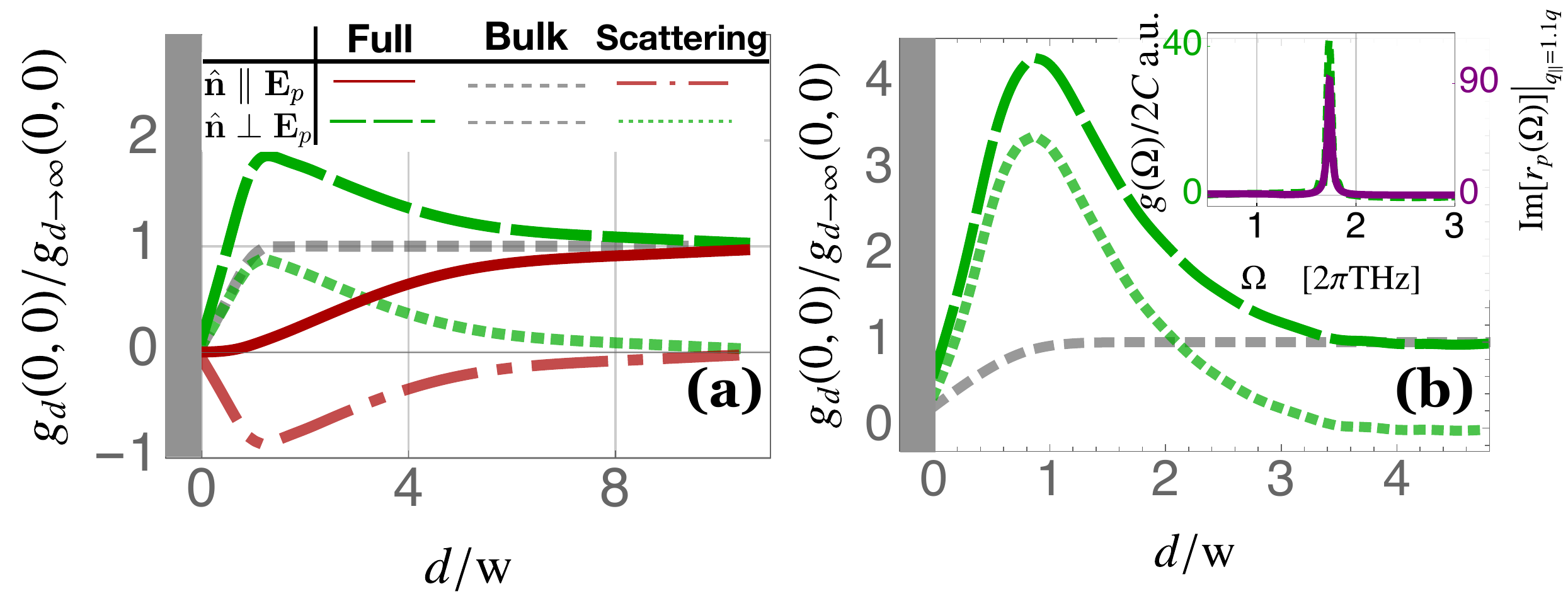}
\caption{\textbf{EOS signal near a plate.} (a) EOS signal as a function of the distance $d$ between the beam center and a perfectly reflecting plate for two different plate orientations: Either the normal of the surface $\hat{n}$ is parallel or orthogonal to the polarization direction of the laser pulses. (b) Same quantity for a single orientation of a plate with Drude-Lorentz model dielectric function. The inset displays the spectrum of the signal alongside that of the imaginary part of the Drude-Lorentz model reflection coefficient at $q_\parallel = 1.1 q $, showing their coinciding peaks. The results are normalised to the value $g^{(0)}_{d\to \infty}$ found when the plate is removed, and we use $L=1\,$mm. }\label{fig:OnePlate}
\end{figure}
We  focus our attention on the changes these objects induce in the quantum vacuum correlations, and therefore neglect their influence on the laser pulses (aside from obscuring part of the beams). This can be justified by assuming that the reflection coefficients of the plate and the cavity are close to zero in the frequency range of the laser, but different from zero for the resolved frequencies of the vacuum field. Inserting the full Green's tensor into Eq.~\eqref{eq:S2ofFilter} we get two contributions, one stemming from the bulk $\tens{G}^{(0)}$ and one from the scattering part $\tens{G}^{(1)}$. The latter describes the change of the correlation function due to the presence of the macroscopic plate(s). Neglecting absorption effects inside the nonlinear crystal (e.g. $\varepsilon(\Omega)$ real-valued) and applying the `laser paraxial' approximation suitable in the parameter range considered here \cite{lindel2020theory,PRAlong,SUPP} the corresponding bulk ($j=0$) and scattering ($j=1$) contributions take the form \cite{SUPP}
 \begin{multline} \label{eq:Signal}
g^{(j)}(\delta t, \delta \vec{r}_\parallel )  = C \int\limits_0^\infty\dif \Omega \, \mathrm{cos}(\Omega \delta t) E_\mathrm{vac}^{2}(\Omega) \int \frac{\dif^2 q_{\parallel}}{4\pi q^2} \\
 \times   R^{2}(\vec{q}) \mathrm{Re}[p^{(j)}(\vec{q},\delta \vec{r}_\parallel)O^{(j)}(\vec{q},\delta \vec{r}_\parallel)].
 \end{multline}
The full signal is given by $g= g^{(0)} +g^{(1)}$. In Eq.~\eqref{eq:Signal}, $\sqrt{C} \propto \chi^{(2)} L I$ ($L$: length of the crystal, $\chi^{(2)}$: nonlinear susceptibility, $I$: intensity of the laser pulses) determines the sampling efficiency \cite{Benea-Chelmus2019} and $E_\mathrm{vac}^{2}(\Omega)$ gives the strength of the vacuum fluctuations in a bulk crystal at frequency $\Omega$. 
The integral over the parallel wave vector of the vacuum field $\vec{q}_\parallel$ ($q = n(\Omega)\Omega/c = \sqrt{q_\perp^2+q_\parallel^2}$) describes the propagation of the virtual photon from one laser pulse to the other. Here, $O^{(j)}(\vec{q},\delta \vec{r}_\parallel)$ accounts for the obscuring of the beam due to the presence of the plate. Most importantly, this integral contains the propagation factor $p^{(j)}(\vec{q},\delta \vec{r}_\parallel)$ which for the bulk contribution is simply given by $p^{(0)} = q_{\parallel}^2/(q_\perp q)\me^{\mi \vec{q} \cdot \delta\vec{r}_\parallel}$, whereas $p^{(1)}$ depends on the chosen geometry of the attached plate(s). For a plate at $x=-d$ for example one finds $p^{(1)} = p^{(0)} \me^{2\mi q_x d}R_p$, where $R_p$ is the p-polarised Fresnel reflection coefficient. The additional factor in $p^{(1)}$ compared to $p^{(0)}$ accounts for the additional propagation of an exchanged virtual photon to the plate and back, see Fig.~\ref{fig:Timedomain}~(a), and further discussion below.  Similar expressions can be found for other geometries and a list of all propagation factors $p^{(1)}$ considered here is found in the Supplementary Material \cite{SUPP}. Furthermore, the integral in Eq.~\eqref{eq:Signal} contains the known response function $R(\vec{q}) = \me^{-(q_x^2+q_y^2)\beamwaist^2/8}\mathrm{sinc}\{L[n_g(\Omega/c)-q_z]/2\} f(\Omega)$ \cite{lindel2020theory,PRAlong} accounting for the averaging over vacuum modes inside the finite lateral pulse profile, phase-matching and which contains the spectral autocorrelation function $f(\Omega) = \me^{-\pi \Omega^2\Delta t^2/4 }$.

\paragraph{Observing the Purcell effect} 
We start by considering a reflecting plate attached to the nonlinear crystal in either the $x=-d$ or $y=-d$ planes, which is thus parallel to the propagation direction of the laser pulses offset by a distance $d$, see inset in the bottom right of Fig.~\ref{fig:SetupWhat}~(a). The contribution to the EOS signal with `coincident' pulses ($\delta t = \delta \vec{r}_\parallel = 0$) is shown in Fig.~\ref{fig:OnePlate} as a function of $d$. \\
First, in Fig.~\ref{fig:OnePlate}~(a), we assume a perfectly reflecting plate described by reflection coefficients for $p$ and $s$ polarised waves $R_p = 1$ and $R_s=-1$, respectively. 
The electro-optic sampling signal changes with the beam-plate distance as a result of competition between two effects. On the one hand, the signal decreases when the beam is closer to the plate, since a larger fraction of the beam becomes obscured by it. On the other hand, the effects of reflection upon the vacuum field start contributing significantly at a distance of approximately $d = 4\beamWaist$. The opposite signs of these additional, plate-induced contributions (`scattering contributions') for the cases where the plate is in the $x=-d$ or $y=-d$ plane can be understood by realising that they arise from image fluctuations: in the former (latter) geometry the image-fluctuations are parallel (antiparallel) with respect to the $x$-polarised fluctuations the laser pulse singles out. This leads to same (opposite) signs of the image fluctuating field compared to the bare fluctuations and thus to an enhancement (reduction) of the total fluctuating field. In both cases, the influence of the scattering contributions and thus of the plate-induced changes upon the quantum-vacuum correlations is clearly visible in the predicted full EOS signal.



A second model for the optical response of the plate is a Drude-Lorentz model permittivity defined by
$\varepsilon( \Omega ) = \varepsilon_\infty \left[ 1+ {\omega_p^2}/(\Omega^2-\omega_\textrm{c}^2 + \mi \Omega \Gamma)     \right]$
with results shown in  Fig.~\ref{fig:OnePlate}~(b) for parameters $\varepsilon_\infty = 8$, $\omega_p = 0.86\times 2\pi\,\textrm{THz}$, $\omega_\textrm{c} = 0.04 \times 2\pi\,\textrm{THz}$ and $\Gamma= 0.056\times 2\pi\,\textrm{THz}$.
These parameters are chosen such that the material's surface-plasmon polariton resonance coincides with the frequencies that the filter function picks out from the vacuum. Consequently there is a peak in the imaginary part of the Fresnel reflection coefficient $R_p$ for $p$-polarized light as can be seen in Fig.~\ref{fig:OnePlate}c, which corresponds to modes evanescent at the interface between the plate and the crystal. These evanescent modes dominate the spectrum of the vacuum's contribution to the variance and lead to an increase of the EOS signal by up to a factor of $5.5$ when the beam gets close to the surface, cf. Fig.~\ref{fig:OnePlate}~(b).  

We thus have revealed in which way one can identify the Purcell effect, i.e. the changes of the medium-assisted quantum vacuum due to the coupling of the electromagnetic field to media, in the statistics of EOS experiments. This makes the latter a suitable tool for an in-depth study of the sculpted quantum vacuum in different optical environments. \\

\paragraph{Time domain perspective}

 Thus far we have restricted the discussion to a frequency domain picture: the modes of the quantum-vacuum fluctuating at different frequencies are accessed by averaging them over the finite space-time volume of the laser pulses, cf. Eq.~\eqref{eq:S2ofFilter}. Revealing the microscopic processes involved leads to a complementary time-domain picture. As we discuss in more detail in the supplementary material \cite{SUPP}, the basic mechanism leading to the EOS signal in Eq.~\eqref{eq:S2ofFilter} are two successive, nonlinear processes which are correlated via the quantum vacuum: in the first process a (virtual) photon is generated into the vacuum mode at position $\vec{r}$ [$\propto \hat{E}_{\mathrm{vac},x}^\dagger(\vec{r})$] which is annihilated in a second process at position $\vec{r}^{\prime}$  [$\propto \hat{E}_{\mathrm{vac},x}(\vec{r}^\prime)$]. One example of two such correlated processes is displayed in Fig.~\ref{fig:Timedomain}~(a) and is given by spontaneous parametric down-conversion and subsequent sum-frequency generation. The two photons which arise from this process are hence correlated and this correlation, which is proportional to the vacuum correlation function between the points $\vec{r}$ and $\vec{r}^{\prime}$, is measured in EOS experiments of the quantum vacuum. On a microscopic level the observed process is hence an exchange of a virtual photon between two points within the nonlinear medium.\\
The emerging time-domain picture of these experiments goes beyond their previous interpretation as a means to measure static, pre-existing vacuum-noise which can be squeezed or shaped by e.g. a cavity. Rather, the signal is interpreted as arising from the propagation of a virtual photon, which thus should experience retardation effects. To illustrate this, we study the signal as a function of the time delay between the pulses $\delta t$ using different values for the beam separations $\delta y$, see Fig.~\ref{fig:Timedomain}~(b). In Fig.~\ref{fig:OnePlate}~(b) one sees that when $\delta y $ becomes greater than the beam waist $\beamwaist$ we find maximal correlations for space-time regions which are shifted also in time by $\delta y c_n$ as dictated by the finite velocity of the virtual photons. This is a clear signature of retardation effects in the quantum-vacuum correlations which also underlie all vacuum induced phenomena such as the Purcell effect and e.g. leads to  the causal behaviour of Van der Waals forces between atoms mediated by such virtual photons \cite{rizzuto2004dynamical,passante2006causality}.

\paragraph{Observing the Purcell effect in time domain} 

In the virtual-photon picture of the quantum vacuum outlined in the last section, the changes of the quantum vacuum due to the presence of additional surfaces can be understood as follows: the virtual photon can not only propagate directly from $\vec{r}$ to $\vec{r}^\prime$, but also propagate along a path which includes reflections from boundaries, see right hand side of Fig.~\ref{fig:Timedomain}~(a). The former (latter) is described by the bulk (scattering) Green's tensor $\tens{G}^{(0)}$ ($\tens{G}^{(1)}$). Again, the dynamics of this process are not arbitrarily fast and, thus, in order to resolve it, the laser pulses which accesses the quantum vacuum must scan the quantum correlations for a time interval long enough such that (multiple) reflections can occur. This leads to a type of time-frequency uncertainty relation.
To show 
this dynamical aspect of the medium-induced quantum vacuum, we consider the effect of reflections from the front and back surfaces of the electro-optic crystal: these already constitute a planar cavity structure, see inset in the bottom right of Fig.~\ref{fig:SetupWhat}~(a). The electro-optic signal can be computed via Eq.~\eqref{eq:Signal} as before with $O^{(0)} = O^{(1)}= 1$ and $p^{(j)}$ given in the supplementary material \cite{SUPP}. By definition, the bulk contribution remains unaffected by the surfaces, whereas for the scattering contribution the phase-matching condition in the response function selects only certain propagation paths. In order to be phase-matched, the virtual photon before and after reflection must propagate in the same direction as one of the two pulses such that it can be `picked up' by that laser pulse. This means, for laser pulses propagating in the $z$-direction, only virtual photons which have been reflected at least twice can be phase-matched, see illustrations in Fig.~\ref{fig:Timedomain}~(c). 

\begin{figure}
  \centering
\includegraphics[width = 1.\columnwidth]{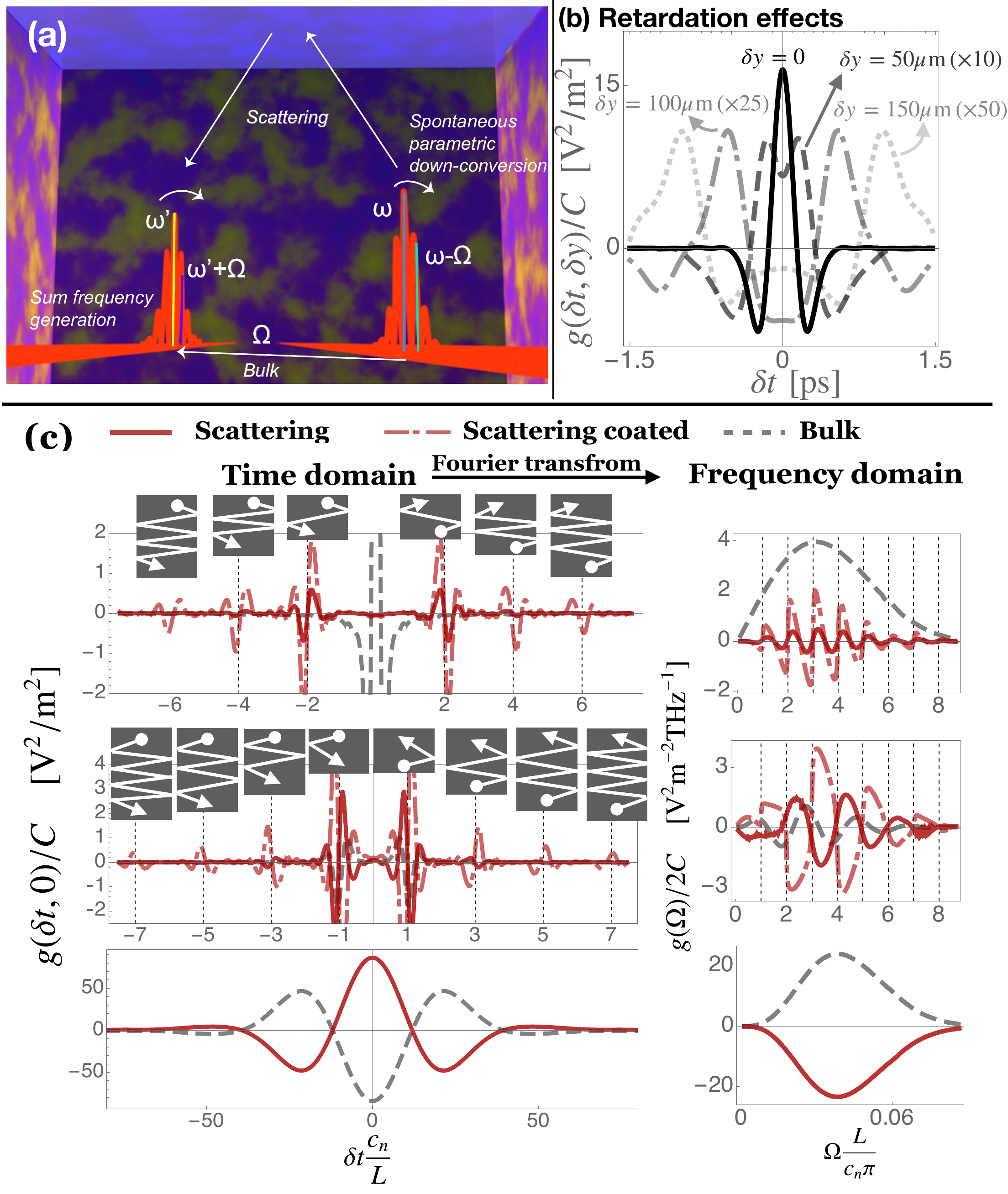}
\caption{\textbf{Time and frequency domain EOS signals inside a cavity} (a) Relevant processes resulting in the EOS signal. (b) EOS signal as a function of the time delay $\delta t$ and for different spatial offsets $\delta y$ between the pulses. (c) The bulk and scattering contributions to the EOS signal $g(\delta t)/C$ as well as its spectrum $g(\Omega)/2C$ are shown. In the first and third row both pulses are propagating into the $z$-direction, whereas in the second row the delayed pulse propagates into the $-z$ direction. We consider coated crystal front and back sides such that $R_p = -R_s =0.95$ (`scattering coated') as well as Fresnel reflection coefficients (`scattering') \cite{SUPP}.} 
\label{fig:Timedomain}
\end{figure}
The resulting signal as a function of $\delta t$ is shown in the first row of Fig.~\ref{fig:Timedomain}~(c). One finds that the scattering contribution is very small for $\delta t= 0$, since the pulses have already left the crystal (cavity) before the virtual photons have been reflected at least twice (this takes $2L/c_n  = 2.1\,\mathrm{ps}$, pulse duration $\Delta t = 80\,$fs). This is again a clear signature of retardation effects in the vacuum correlation function. However, if the delay between the pulses $\delta t $ is a multiple of $2L/c_n$ the second pulse arrives at the crystal precisely when the virtual photon generated by the first pulse has been reflected an even number of times. Thus the peaks for even values of $\delta t c_n /L$ in Fig.~\ref{fig:Timedomain}~(c) display its propagation in time domain. Fourier transforming the signal with respect to $\delta t$ one can obtain the signal's spectrum $g(\Omega)$ [$g(\delta t= 0) = \int \dif \Omega g(\Omega)$] \cite{PRAlong}, which shows the expected mode structure [see Fig.~\ref{fig:SetupWhat}~(b)] with resonances at multiples of $c_n L /\pi$, compare right hand side of Fig.~\ref{fig:Timedomain}~(c). However, in order to obtain $g(\Omega)$ experimentally one has to perform a measurement of $g(\delta t)$ for a wide range of $\delta t$, i.e. one needs to resolve the correlations arising from different numbers of reflections individually in time. In the case of a single measurement at $\delta t = 0$, the positive and negative contributions in $g^{(1)}(\Omega)$ mostly cancel each other such that the cavity field remains unseen when it is averaged over a single space-time region. This can be seen as a time-frequency uncertainty relation.
              
To further improve the visibility of the cavity contribution we consider two beams propagating in opposite directions, i.e. the first into $z$, the second into $-z$. As a result, the bulk contribution is phase-mismatched and reduced considerably \cite{SUPP}. The scattering contribution is dominated by that stemming from virtual photons which are reflected an odd number of times: they are generated by the first laser pulse propagating into $z$ direction and are `picked up' while propagating into $-z$ direction by the second laser pulse. Since these virtual photons only have to propagate over a shorter distance compared to the configuration in which both pulses propagate into the same direction, this improves the visibility of the cavity modes, cf. second row of Fig.~\ref{fig:Timedomain}~(c). 

Lastly, we consider the case where the durations of the laser pulses are much longer then the time a photon needs to propagate back and forth between the front and back side of the crystal, i.e. $\Delta t \gg L/ c_n$. In this case, multiple reflection can in principle occur but the spectral autocorrelation function restricts the accessed quantum vacuum to frequencies much shorter than the lowest resonant mode, i.e. $f (\Omega) = 0$ for all $\Omega> c_n \pi/L.$ Hence, all the accessed modes interfere destructively so that in this case the vacuum field is completely suppressed such that $g(0,0) \approx 0$, compare lowest row of Fig.~\ref{fig:Timedomain}~(c). 


\paragraph{Conclusion}

In  this work we have shown how electro-optic sampling can be exploited to measure environment-induced changes of the electromagnetic vacuum fluctuations (i.e. the Purcell effect) in the frequency and time domains. Interpreting vacuum correlations as arising from the exchange of virtual photons leads to a time domain picture of how vacuum correlations evolve which reveals retardation effects and a time-frequency uncertainty relation. This dynamical picture might allow one to reveal other space-time properties of the quantum vacuum in the future \cite{Biswas, Valentini, Reznik1, Martinez}. It also forges a missing link between EOS and well established quantum-vacuum effects which can also be seen as arising from the exchange of virtual photons such as e.g. van der Waals forces, Casimir forces or resonant energy transfer and the influence of different optical environments upon them. Future extensions might include studying the vacuum field in more complex geometries such as plasmonic cavities \cite{benea2020electro}.

 \begin{acknowledgments}
The authors thank J\'{e}r\^{o}me Faist, Alexa Herter, Stephen Barnett, Denis Seletskiy, Guido Burkard and Alfred Leitenstorfer for fruitful discussions. R.B. acknowledges financial support by the Alexander von Humboldt Foundation, S.Y.B. thanks the Deutsche Forschungsgemeinschaft (grant BU 1803/3-1476). F.L. acknowledges support from the Studienstiftung des deutschen Volkes.
\end{acknowledgments}


%

\newpage
\clearpage
\newpage

\begin{widetext}

\begin{center}
\large{\textbf{Supplemental material: `Probing the Purcell effect without radiative decay: Lessons in the frequency and time domains'}}
\end{center}

\normalsize

\section{Green's tensor} \label{sec:Green}

The Green's tensor of the vector Helmholtz equation is defined via \cite{Buhmann2012a2}
\begin{align} \label{eq:GreensTensorDef}
\left( \nabla \times \nabla \times - \frac{\Omega^2}{c^2} \varepsilon(\vec{r},\Omega)  \right) \tens{G}(\vec{r},\vec{r}^\prime,\Omega) = \Greektens{\delta}(\vec{r} - \vec{r}^\prime ),
\end{align}
with the boundary condition \mbox{$\tens{G}(\vec{r},\vec{r}^\prime,\Omega) \to 0 $} for \mbox{$|\vec{r} - \vec{r}^\prime| \to \infty$}. $\tens{G}$ can be subdivided into its bulk ($\tens{G}^{(0)}$) and scattering ($\tens{G}^{(1)}$) components such that $\tens{G} = \tens{G}^{(0)} + \tens{G}^{(1)}$. \\

\subsection{Bulk Green's tensor}

The bulk Green's tensor solves Eq.~\eqref{eq:GreensTensorDef} for an isotropic permittivity, i.e. \mbox{$\epsilon(\vec{r},\Omega) =\epsilon(\Omega)$}. In a $(2+1)$-dimensional Weyl decomposition relative to a plane whose normal direction is denoted by $\rPerp$ it reads: \cite{Buhmann2012a2}
\begin{multline} \label{eqapp:BulkGreensTensor} 
\tens{G}^{(0)}(\vec{r},\vec{r}^\prime, \Omega) = - \frac{1}{4\pi^2 k^2(\omega)} \int \dif^2 k_\parallel \, \frac{\me^{i\vec{k}_\parallel \cdot (\vec{r}-\vec{r}^\prime)}}{\kPerp}\delta(\rPerp - \rPerp^\prime) \vec{e}_\perp \vec{e}_\perp
\\ + \frac{i}{8\pi^2} \int \dif^2 k_\parallel \frac{\me^{i\vec{k}_\parallel \cdot (\vec{r}-\vec{r}^\prime)}}{\kPerp} \!\!\! \sum_{\sigma=s,p} \! \left[ \vec{e}_{\sigma+}\vec{e}_{\sigma+} \me^{i \kPerp(\rPerp-\rPerp^\prime)} \theta(\rPerp \!\! - \rPerp^\prime) \right. \\
\left. +\vec{e}_{\sigma-}\vec{e}_{\sigma-} \me^{-i \kPerp(\rPerp-\rPerp^\prime)} \theta(\rPerp^\prime \! \!- \rPerp)\right].
\end{multline}
Here, we have defined the wave vector $k =\sqrt{\epsilon(\Omega)} \Omega/c$, which can be split into perpendicular ($\kPerp$) and parallel ($k_\parallel = |\B{k}_\parallel|$) components. Note that \mbox{$\kPerp = \kPerp(k_\parallel, \Omega)=\sqrt{k^2-k_\parallel^2}$} with \mbox{$\textrm{Im}[\kPerp]>0$}. The polarization vectors $\vec{e}_{\sigma\pm}$, $\sigma = s,p$, are defined via
\begin{align} \label{eq:GreensTensorPolarizationVectors}
\vec{e}_{s\pm}(\vec{k}_\parallel) & = \vec{e}_{k_{\parallel}} \times \vec{e}_\perp  ,\\
\vec{e}_{p\pm}(\vec{k}_\parallel)  & = \frac{1}{k}(k_{\parallel} \vec{e}_{\perp} \mp \kPerp \vec{e}_{k_{\parallel}}).
\end{align}

\subsection{Scattering Green's tensor}

The scattering part of the Green's tensor depends on the geometry under consideration. Here, we are firstly interested in the geometry of a plate attached to the nonlinear crystal. Note, that we here neglect reflections from the surfaces of the crystal. We assume that the plate is thick enough such that it can be approximated by a semi-infinite half space with refractive index $n^\prime(\Omega)$ whose interface is placed at $\rPerp < -d$, with $\rPerp$ being the plate's normal vector. For the refractive index used to characterize the attached plate, see Sec.~\ref{sec:Param} of this Supplementary Matrial. For $\rPerp > -d$ we thus find the nonlinear crystal with refractive index $n(\Omega)$. In part b of the main text, we consider precisely this configuration with $\rPerp = x$ or $\rPerp = y$. The scattering part of the Green's tensor inside the nonlinear crystal ($\rPerp, r_\perp^\prime > - d $) reads \cite{Buhmann2012a2}
\begin{align} \label{eqapp:ScatteringGreensTensorPlate}
\tens{G}^{(1)}(\vec{r},\vec{r}^\prime, \Omega) = \frac{i}{8\pi^2} \int \dif^2 k_{\parallel} \frac{\me^{i\vec{k}_{\parallel} \cdot (\vec{r}-\vec{r}^\prime) + \mi \kPerp(\rPerp + \rPerp^\prime + 2 d)}}{k_\perp} 
 \sum_{\sigma=s,p} \rSigma \vec{e}_{\sigma +}\vec{e}_{\sigma -}.
\end{align}
As in the main text, $\rSigma$ are the reflection coefficients at the nonlinear crystal/plate interface. In case of perfectly reflecting plates they are given by $R_p = 1$, $R_s = -1$ and in case of a plate with finite permittivity $\epsilon^\prime$ they are the usual Fresnel reflection coefficients which are given by
\begin{align} \label{eq:Refl1}
R_p & = \frac{n^\prime(\Omega) k_{\perp} -  n(\Omega) k_{\perp}^\prime }{n^\prime(\Omega) k_{\perp} +  n(\Omega) k_{\perp}^\prime }, \\ \label{eq:Refl2}
R_s  &  = \frac{ k_{\perp} -  n(\Omega) k_{\perp}^\prime }{ k_{\perp} +  n(\Omega) k_{\perp}^\prime },
\end{align}
where $k_{\perp}^\prime =\sqrt{n^{\prime 2}(\Omega)\Omega^2/c^2 - q_\parallel^2}$. \\
 %
 
The other geometry considered in this work is that where reflections at the surfaces of the crystal are taken into account. This can be done by including the scattering part of the Green's tensor for a configuration where the refractive index for $-L/2< z<L/2$ is given by the one of the nonlinear crystal and otherwise it is defined to be the vacuum refractive index, i.e. $n^\prime(\omega) =1$. For this geometry one finds \cite{Buhmann2012a2}
\begin{multline} \label{eqapp:GreensTensorCavity}
\tens{G}^{(1)}(\vec{r},\vec{r}^\prime, \omega) = \frac{i}{8\pi^2} \int \dif^2 k_{\parallel} \frac{\me^{i\vec{k}_{\parallel} \cdot (\vec{r}-\vec{r}^\prime)}}{k_\perp} 
 \sum_{\sigma=s,p} \\
 \times \left\{ \frac{\rSigma \rSigma \me^{2\mi \kPerp L}}{D_\sigma} \left[ \vec{e}_{\sigma + }\vec{e}_{\sigma +} \me^{\mi \kPerp(\rPerp-\rPerp^\prime)} + \vec{e}_{\sigma - }\vec{e}_{\sigma -} \me^{-\mi \kPerp(\rPerp-\rPerp^\prime)} \right] \right. \\
 \left. + \frac{1}{D_\sigma}  \left[   \vec{e}_{\sigma + }\vec{e}_{\sigma -} \rSigma \me^{\mi \kPerp(\rPerp+\rPerp^\prime+ L )}  +\vec{e}_{\sigma - }\vec{e}_{\sigma +} \rSigma  \me^{\mi \kPerp[ L -\rPerp-\rPerp^\prime]}  \right]           \right\},
\end{multline}
for $\rPerp, \rPerp^\prime \in [-L/2,L/2]$. Note that here we have chosen $\rPerp = z$ and \mbox{$k_\perp = k_z = \sqrt{k - k_x^2 -k_y^2}$} and the terms $D_\sigma  =1- \rSigma \rSigma \me^{2\mi k_\perp L}$ in the denominators account for multiple reflections. In the case that we assume that the crystal's surfaces are coated in order to increase the reflectivity in the THz range, we assume $R_p=  0.95$ and $R_s =-0.95$ and otherwise the reflection coefficients are given by Eqs.~\eqref{eq:Refl1} and \eqref{eq:Refl2} with $n^\prime(\omega) = 1$.

\section{Parameters} \label{sec:Param}

 In this section we give all optical parameters of the nonlinear crystal and its optical surroundings used in the main text to simulate the signal of electro-optic sampling experiments.

The ordinary and group refractive indices of the nonlinear crystal at $\omega_c$ are $n=2.85$ and $n_g=3.2$ as measured in Ref.~\cite{Leitenstorfer19992}. For the nonlinear refractive index we neglect its dispersion and use \cite{Leitenstorfer19992}
\begin{align}\label{42:NonlinearSusDisp}
\chi^{(2)}(\Omega) \approx  \chi^{(2)} =  \frac{n^4(\omega_\textrm{C})\epsilon_0}{2} r_{41},
\end{align}
with $r_{41}=1.17\times 10^{-21}\,\textrm{C}\textrm{V}^{-2}$ \cite{Leitenstorfer19992}. \\

 In the THz range we use the data for $n$ which was measured using time domain spectroscopy in Ref.~\cite{Benea-Chelmus20192}. Its resulting real and imaginary part are shown in Fig.~\ref{fig:Refr}.
\begin{figure}
\includegraphics[scale=0.4]{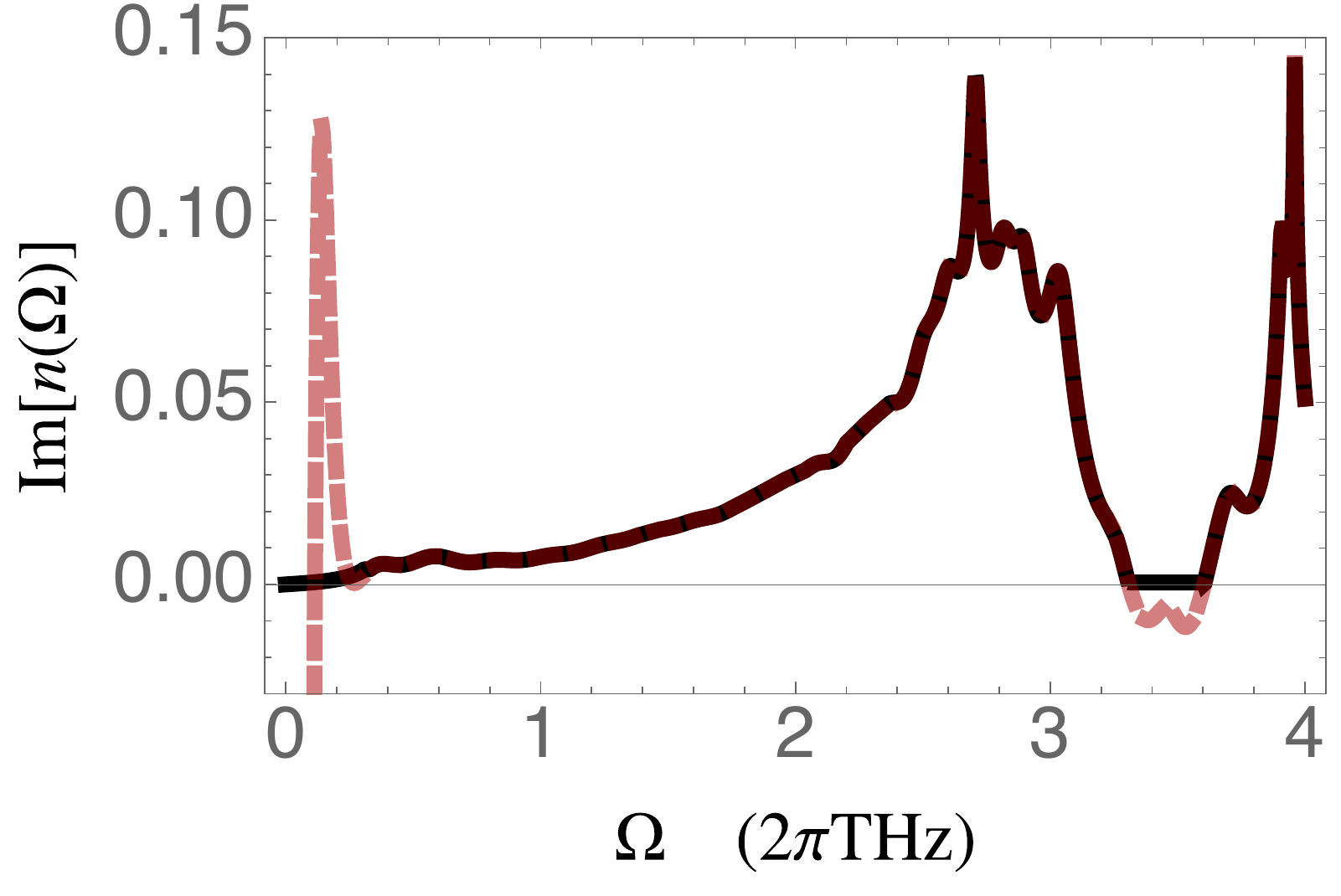}\quad \quad \quad 
\includegraphics[scale=0.4]{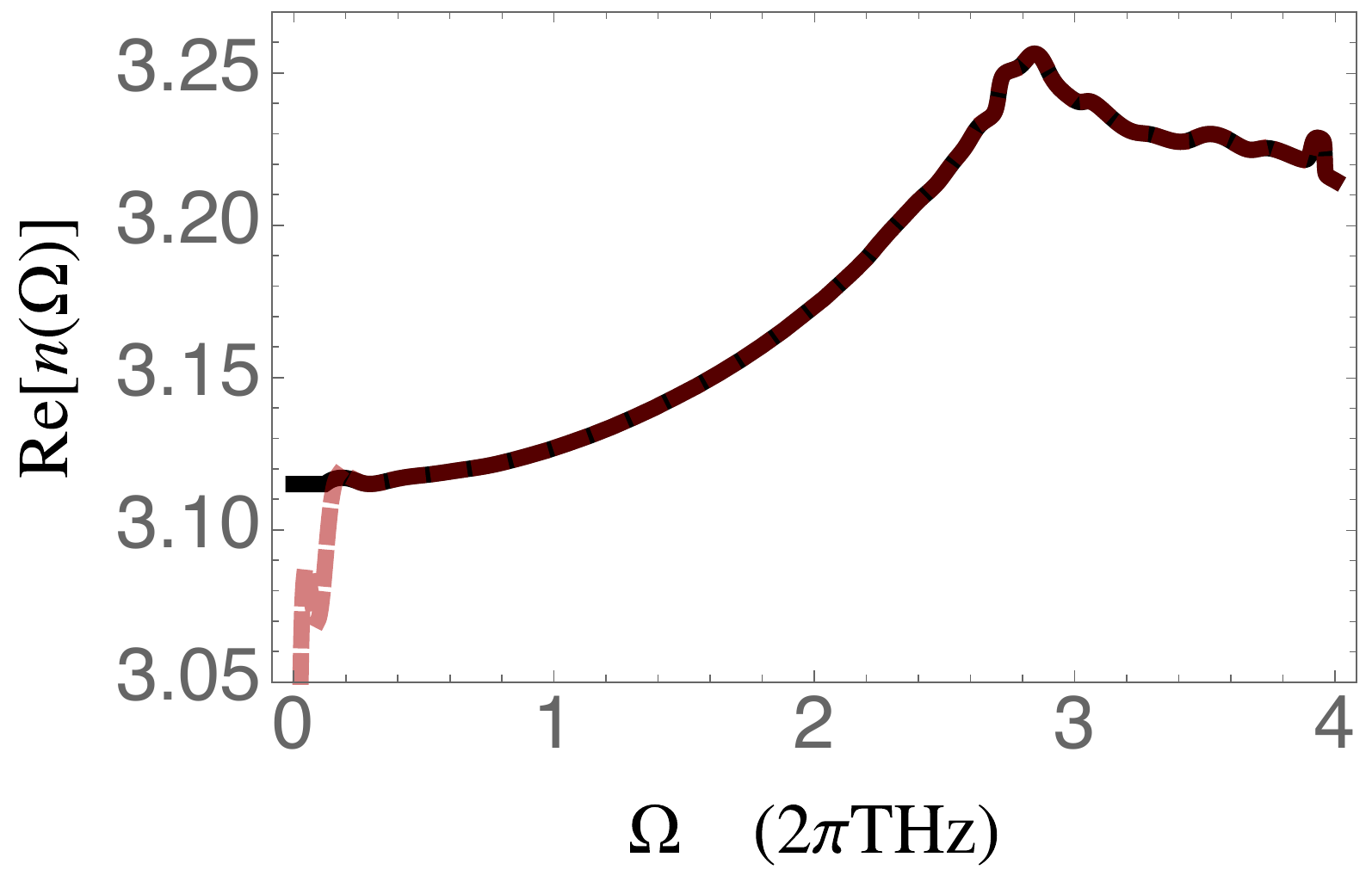} 
\caption{Real and imaginary parts of the refractive index in the THz as measured in Ref.~\cite{Benea-Chelmus20192}. The dashed red lines show an interpolation line of the measured data whereas the black solid line indicates the refractive index used in the simulations. The latter differs from the measured data in frequency ranges where the measured data is not reliable, i.e. for small frequencies and in region where Im$[n(\Omega)] <0$. }
\label{fig:Refr}
\end{figure}  
 
 For the plate attached to the nonlinear crystal which is described by a Drude--Lorentz model (Fig. 2b in the main text), we have used $ n^\prime(\Omega) =  \sqrt{\epsilon^\prime(\Omega)}$ with
\begin{align} \label{eq:PermittivityDrude}
\epsilon^\prime( \Omega ) = \epsilon_\infty \left[ 1+ \frac{\omega_\textrm{p}^2}{\Omega^2-\omega_\textrm{c}^2 + \mi \Omega \Gamma}     \right],
\end{align}
and $ \epsilon_\infty=8$, $\omega_\textrm{p} = 0.86 \times (2\pi)$\,THz, $\omega_\textrm{c} = 0.04 \times (2\pi)$\,THz, $\Gamma= 0.56 \times (2\pi)$\,THz.

\section{Theory of electro-optic sampling in general environments }\label{app:EOS}

In this section we derive analytic expressions for the electro-optic signal in the different geometries considered in the main text. The general approach is always the same and build upon Eqs.~(1) and (2) of the main text: One has to find the filter function as well as the Green's tensor for the different configurations, insert these expressions into Eqs.~(1) and (2) of the main text and solve as many of the resulting integrals as possible.

\subsection{General equations}

As shown in Refs.~\cite{Moskalenko20152,PRAlong2,lindel2020theory2}, the electro-opitc signal is given by
\begin{multline} \label{eq:S2g}
g = \langle:  \mathcal{S}\left\{ \left( 4\pi \epsilon_0 c \int_0^\infty \dif \omega \frac{\eta(\omega)}{\hbar \omega} \int \dif^2 r_\parallel  \left[ \mi \hat{E}_{1,y}^\dagger(\vec{r}_\parallel, \omega) \hat{E}_{1,x}(\vec{r}_\parallel, \omega) + \text{h.c.} \right] \right)\right.  \\\left.
\times \left( 4\pi \epsilon_0 c \int_0^\infty \dif \omega \frac{\eta(\omega)}{\hbar \omega} \int \dif^2 r_\parallel  \left[ \mi \hat{E}_{2,y}^\dagger(\vec{r}_\parallel, \omega) \hat{E}_{2,x}(\vec{r}_\parallel, \omega) + \text{h.c.} \right] \right) \right\} : \rangle .
\end{multline}
Here, $\eta(\omega)$ is the detector efficiency which is assumed to be one over the frequency range of the laser pulses, $\mathcal{S}$ is the symmetrization operator which is defined by its action onto a product of two operators $\hat{O}_{1,2}$: $\mathcal{S} \hat{O}_{1} \hat{O}_{2} = (\hat{O}_{1} \hat{O}_{2} +  \hat{O}_{2} \hat{O}_{1})/2$, $: \dots :$ denotes normal ordering, $\langle \dots \rangle$ means that the ground state expectation value is taken, and $\hat{\vec{E}}_{1,2}$ is the electric field at detector $1$ and $2$, respectively. $\hat{\vec{E}}_{1,2}$ can be perturbatively expanded in terms of the nonlinear susceptibility $\chi^{(2)}$ and consists of the vacuum field $\hat{\vec{E}}_\textrm{vac}$, the classical probe field $1$ ($\vec{E}_\textrm{p,1}$) \textit{or} $2$ ($\vec{E}_\textrm{p,2}$) and the field consisting of the nonlinear mixing of the vacuum field with either $\hat{\vec{E}}_{p,1}$ or $\hat{\vec{E}}_{p,2}$, respectively. This means, the electric field emerging from the crystal consists of the free fields (vacuum and coherent laser pulses) as well as of their mixing via the nonlinear coupling inside the crystal. See Refs.~\cite{PRAlong2,lindel2020theory2} for details.
Allowing for absorption inside the crystal as well as general optical environments it was shown \cite{PRAlong2,lindel2020theory2} that $g$ is given by Eqs.~(1) and (2) of the main text which are repeated here for clarity
\begin{align} \label{eq:S2ofFilter2}
g(\delta t, \delta \vec{r}_\parallel)  & = \int_{V_\textrm{C}} \!\! \dif^3 r \!\! \int_{V_\textrm{C}} \!\! \dif^3 r^\prime\!\! \int_0^\infty \hspace{-0.3cm} \dif \Omega \int_0^{\infty} \hspace{-0.3cm}\dif \Omega^\prime  F(\vec{r},\vec{r}^\prime, \Omega, \Omega^\prime,\delta \vec{r}_\parallel, \delta t) \langle \hat{ E}_\mathrm{vac,x}(\vec{r},\Omega) \hat{E}_\mathrm{vac,x}^\dagger(\vec{r}^\prime,\Omega^\prime) \rangle,  \\ \label{eq:EvacEvac}
\langle \hat{ \vec{E}}_\mathrm{vac}(\vec{r},\Omega) \hat{\vec{E}}_\mathrm{vac}^\dagger(\vec{r}^\prime,\Omega^\prime) \rangle & = \frac{\hbar \Omega^2 }{c^2 \varepsilon_0\pi} \delta (\Omega-\Omega^\prime) \mathrm{Im} \tens{G}(\vec{r},\vec{r}^\prime,\Omega), \\ \label{eq:S2ofFilterAndG}
\Longrightarrow \quad g(\delta t, \delta \vec{r}_\parallel)  & = \frac{\hbar \Omega^2 }{c^2 \varepsilon_0\pi}  \int_{V_\textrm{C}} \!\! \dif^3 r \!\! \int_{V_\textrm{C}} \!\! \dif^3 r^\prime\!\! \int_0^\infty \hspace{-0.3cm} \dif \Omega   F(\vec{r},\vec{r}^\prime, \Omega, \Omega^\prime,\delta \vec{r}_\parallel, \delta t)   \mathrm{Im} \tens{G}(\vec{r},\vec{r}^\prime,\Omega).
\end{align} 
The filter function is given by 
\begin{multline}\label{eq33:FilterOfH}
F(\vec{r},\vec{r}^\prime,\Omega,\Omega^\prime) = \frac{1}{2}\left\{\left[  H_1(\vec{r},\Omega) + H_1^\ast(\vec{r},-\Omega) \right]\left[ H_2(\vec{r}^\prime,-\Omega^\prime) + H_2^\ast(\vec{r}^\prime,\Omega^\prime) \right] \right. \\
\left. + \left[  H_2(\vec{r},\Omega) + H_2^\ast(\vec{r},-\Omega) \right]\left[ H_1(\vec{r}^\prime,-\Omega^\prime) 
 + H_1^\ast(\vec{r}^\prime,\Omega^\prime) \right] \right\},
\end{multline}
with
\begin{align}\label{eq:hfunction}
H_i(\vec{r}^\prime,\Omega)  = -8 \pi \mi  c \epsilon_0  \chi^{(2)}(\Omega) \mu_0 \int\limits_0^\infty \dif\omega \frac{\eta(\omega) \sqrt{\epsilon(\omega)} \omega}{\hbar}  \int \dif^2 r_\parallel E_{\textrm{p},i,y}^\ast(\vec{r}_\parallel,\omega) \textsf{G}_{xx} (\vec{r}_\parallel,\vec{r}^\prime, \omega)E_{\textrm{p},i,y}(\vec{r}^\prime, \omega-\Omega).
\end{align}
To further simplify this expression, we insert the Gaussian laser pulse defined in the main text for $\vec{E}_{p,i}$ into Eq.~\eqref{eq:hfunction}, i.e. we use $E_\mathrm{p,i}(\vec{r},t)  = \int \dif \omega  E_\mathrm{p,i}(\vec{r},\omega) \me^{\mi \omega t}$ with
\begin{align}
\vec{E}_\mathrm{p,1}(\vec{r},\omega) &=E_\mathrm{p}(\omega)\sqrt{\frac{2}{\pi \beamwaist^2}}\me^{-\vec{r}_\parallel^2/\beamwaist^2} \me^{\mi k z} \vec{e}_y, \\ \label{eq:E2}
\vec{E}_\mathrm{p,2}(\vec{r},\omega) & =E_\mathrm{p}(\omega)\sqrt{\frac{2}{\pi \beamwaist^2}}\me^{-(\vec{r}_\parallel - \delta\vec{ r}_\parallel)^2/\beamwaist^2} \me^{\mi k z} \me^{\mi \omega  \delta t } \vec{e}_y, \\
E_\mathrm{p}(\omega) & = \sqrt{\frac{\Delta t}{2}} \me^{-\pi(\omega-\omega_\mathrm{c})^2\Delta t^2/2} .
\end{align}
Note, that here $\vec{r}_\parallel = (x,y,0)^T$, and $k = n(\omega) \omega/c$. 
Also we insert $ \textsf{G}_{xx}(\omega) =  \textsf{G}^{(0)}_{xx}(\omega) $ with $\tens{G}^{(0)} $ given in Eq.~\eqref{eqapp:BulkGreensTensor} into Eqs.~\eqref{eq33:FilterOfH} and \eqref{eq:hfunction}. This is justified since throughout this work we assume that the optical environments do not affect the near-infrared laser pulses apart from obscuring them as discussed in the main text. Neglecting absorption in the frequency range of the laser pulses and applying the laser-paraxial approximation introduced in Refs.~\cite{PRAlong2,lindel2020theory2} one finds
  \begin{multline} \label{eqapp:FgeneralisedPPGRI}
F(\vec{r},\vec{r}^\prime,\Omega) = 2 \left( \frac{2 \chi^{(2)}\mu_0 c N \omega_p f(\Omega)}{\beamWaist^2 n(\omega_c)} \right)^2 \me^{-\mi n_g \Omega(z-z^\prime)/c} \\
\times \left[ \me^{-2\left[\vec{r}_\parallel^2 + (\vec{r}_\parallel^\prime + \delta\vec{r}_\parallel)^2 \right]/\beamWaist^2} \me^{\mi \Omega \delta t} +  \me^{-2\left[\vec{r}_\parallel^{\prime 2} + (\vec{r}_\parallel+ \delta\vec{r}_\parallel)^2 \right]/\beamWaist^2} \me^{-\mi \Omega \delta t} \right].
\end{multline}
Here, $n_g$ is the group refractive index at the central frequency $\omega_c$ of the laser pulse and $\omega_p$ and $f(\Omega)$ are the averaged detected frequency and the spectral autocorrelation function, respectively, given by
\begin{align}
\omega_p & =  \frac{\int_0^\infty \dif \omega \eta(\omega) E^2_{p}(\omega) }{ \int_0^\infty \dif \omega \frac{\eta(\omega)}{\omega}  E_p^2 (\omega)} ,\\
f(\Omega)& = \frac{ \int_0^\infty \dif \omega \left[ E_\textrm{p}(\omega) E_\textrm{p}(\omega + \Omega) + E_\textrm{p}(\omega) E_\textrm{p}(\omega - \Omega)  \right]}{ 2\int_0^\infty \dif \omega \eta(\omega) E_\textrm{p}(\omega)^2} \approx \me^{-\pi \Omega^2 \Delta t^2/4}.
\end{align}
After having obtained the filter function, as well as the Green's tensor for the different geometries (see Sec.~\ref{sec:Green}) we have all the ingredients needed in order to calculate the electro-optic sampling signals in the different configurations using Eqs.~\eqref{eq:S2ofFilter2} and \eqref{eq:EvacEvac} which is done in the next two sections.

\subsection{Plate with different orientations parallel to the propagation direction of the laser pulses}

Here, we consider a plate which is parallel to the propagation direction of the laser pulses with two different orientations, i.e. in the $x<-d$ and $y<-d$ half spaces. We neglect absorption inside the nonlinear crystal by assuming $\epsilon(\Omega)$ to be real. One can split the signal $g$ in Eq.~\eqref{eq:S2ofFilterAndG} into contributions stemming from the bulk and the scattering Green's tensors. Note, that although the bulk Green's tensor inside the nonlinear crystal is unaffected by the additional optical plates considered here, its contribution to the EOS signal still changes due to the fact that the laser pulses might get obscured by the plate whenever the pulse/plate distance becomes comparable to the beam waist. This is included in Eq.~\eqref{eq:S2ofFilterAndG} by the restriction of the spatial integrals to the volume of the crystal $V_\textrm{C}$. These integrals become $\int_{V_\textrm{C}} \!\! \dif^3 r = \int_{-L/2}^{L/2} \dif z \int_{-\infty}^{\infty}\dif y\int_{-d}^\infty \dif x $ and $\int_{V_\textrm{C}} \!\! \dif^3 r = \int_{-L/2}^{L/2} \dif z \int_{-\infty}^\infty \dif y \to \int_{-d}^\infty \dif y \int_{-\infty}^{\infty}\dif x $ for the two different orientations of the plate considered here, respectively. Inserting the bulk or scattering Green's tensors in Eqs.~\eqref{eqapp:BulkGreensTensor} and \eqref{eqapp:ScatteringGreensTensorPlate} with $r_\perp$ chosen perpendicular to the applied surface and the filter function in Eq.~\eqref{eqapp:FgeneralisedPPGRI} into Eq.~\eqref{eq:S2ofFilterAndG} one finds after a tedious but straight forward calculation that the signal can always be brought into the form
 \begin{align} \label{eq:Signal2}
g^{(j)}(\delta t, \delta \vec{r}_\parallel )  = C \int\limits_0^\infty\dif \Omega \, \mathrm{cos}(\Omega \delta t) E_\mathrm{vac}^{2}(\Omega) \int \frac{\dif^2 q_{\parallel}}{4\pi q^2}  R^{2}(\vec{q}) \mathrm{Re}[p^{(j)}(\vec{q},\delta \vec{r}_\parallel)O^{(j)}(\vec{q},\delta \vec{r}_\parallel)].
 \end{align}
 This expression is identical to Eq.~(3) of the main text. Here, $\sqrt{C} = 2 \chi^{(2)} L \omega_p N(d,\delta \vec{r}_\parallel)/n \epsilon_0 c$ with $N(d,\delta \vec{r}_\parallel)$ being the total number of detected photons given by
 \begin{align}
N^2(d,\delta \vec{r}_\parallel) & = \frac{1}{4} \left( 1 + \mathrm{Erf}\left[ \frac{\sqrt{2}d}{\beamwaist} \right] \right) \left( 1 + \mathrm{Erf}\left[ \frac{\sqrt{2}(d+ \delta \vec{r}_\parallel \cdot \hat{\vec{n}})}{\beamwaist} \right] \right) N^2 ,\\
 N & = \frac{4\pi \epsilon_0 c n(\omega_c)}{\hbar} \int_0^\infty \dif \omega  \frac{\eta(\omega)}{\omega} E_p^2(\omega).
 \end{align}
$N$ is the total number of detected photons without the plate obscuring parts of the laser pulses, and \mbox{$\mathrm{Erf}[x] = (2/\sqrt{\pi}) \int_0^x \dif t \,\me^{-t^2}$} is the error function. $E_\textrm{vac}^2(\Omega)$ is the coincidence limit of the bulk two-point correlation function of the electric field operator neglecting absorption effects and it reads
\begin{align}
E_\textrm{vac}^2(\Omega)  = \frac{\hbar \mathrm{Re}[n(\Omega)] \Omega^3}{2\epsilon_0 \pi^2 c^3}.
\end{align}
Also, note that the wave vector $\vec{q}$ has been split into a component which is parallel to the surface of the attached plate ($q_\parallel$) and one which is perpendicular to it ($q_\perp = \sqrt{q^2-q_\parallel^2}$), with $q = n(\Omega)\Omega/c$. Finally, as also stated in the main text, the response function $R(\vec{q})$ is given by
\begin{align} \label{eq:Response}
R^2(\vec{q}) & = \me^{-(q_x^2+q_y^2)\beamwaist^2/4} \left\{   \mathrm{sinc}^2\left[ \frac{L}{2} \left(n_g \frac{\Omega}{c}-q_z\right)\right] + \mathrm{sinc}^2\left[ \frac{L}{2} \left(n_g \frac{\Omega}{c}+q_z\right)\right] \right\} f^2(\Omega)\\ \label{eq:Response2}
& \approx \me^{-(q_x^2+q_y^2)\beamwaist^2/4}\mathrm{sinc}^2\left[ \frac{L}{2} \left(n_g \frac{\Omega}{c}-q_z\right)\right] f^2(\Omega).
\end{align}
Here, we restricted the response in the second row to the phase-matched contribution only. Note, that this approximation is only a good approximation for propagating modes with $q_\parallel < \mathrm{Re}[q]$. It thus is used to calculate the EOS signal in Fig.~2 (a) but not in Fig.~2 (b), since in the latter case the main contribution to the scattering part of the signal stems from evanescent modes with $q_\parallel > \mathrm{Re}[q]$. \\
The obscuring factors $O^{(j)}(\vec{q},\delta \vec{r}_\parallel,d)$ in case of a plate in $x<-d$ are given by
\begin{align} \label{eq:Obs1}
O^{(0)}(\vec{q},\delta \vec{r}_\parallel) & =\frac{1}{4} \left( 1 + \mathrm{Erf}\left[ \frac{\sqrt{2} d}{\beamwaist} + \mi \frac{\mi q_x \beamwaist}{2\sqrt{2}}   \right] \right)\left( 1 + \mathrm{Erf}\left[ \frac{\sqrt{2} (d + \delta x)}{\beamwaist} - \mi \frac{\mi q_x \beamwaist}{2\sqrt{2}}   \right] \right), \\ \label{eq:Obs2}
O^{(1)}(\vec{q},\delta \vec{r}_\parallel) & =\frac{1}{4} \left( 1 + \mathrm{Erf}\left[ \frac{\sqrt{2} d}{\beamwaist} + \mi \frac{\mi q_x \beamwaist}{2\sqrt{2}}   \right] \right)\left( 1 + \mathrm{Erf}\left[ \frac{\sqrt{2} (d + \delta x)}{\beamwaist} + \mi \frac{\mi q_x \beamwaist}{2\sqrt{2}}   \right] \right).
\end{align}
The expressions for $O^{(j)}(\vec{q},\delta \vec{r}_\parallel)$ in case of a plate in $y<-d$ can be obtained from Eqs.~\eqref{eq:Obs1} and \eqref{eq:Obs2} by replacing $q_x \leftrightarrow q_y$ and $\delta x \leftrightarrow \delta y$.\\
Finally, the propagation factor $p^{(j)}$ differs for the bulk and scattering contribution as well as for the different geometries (plates in $x<-d$ or $y<-d$) and reads for the different configurations: \\

\centerline{
\begin{tabular}{l | l  l}
 & plate in $x<-d$, $q_\perp = q_x$ & plate in $y<-d$, $q_\perp = q_y$ \\ \specialrule{.2em}{.1em}{.15em} 
 $p^{(0)}(\vec{q},\delta \vec{r}_\parallel)$\hspace{0.3cm} & $\frac{q_\parallel^2}{q_\perp q} \me^{\mi \delta \vec{r}_\parallel  \cdot \vec{q}} $ & $ \frac{q}{q_\perp }\left(1- \frac{q_\parallel^2}{q^2} \right) \me^{\mi \delta \vec{r}_\parallel \cdot \vec{q}} $ \\ \specialrule{.05em}{.1em}{.1em} 
  $p^{(1)}(\vec{q},\delta \vec{r}_\parallel)$ & $\frac{q_\parallel^2}{q_\perp q} \me^{2\mi d q_\perp} R_p  \me^{\mi \delta \vec{r}_\parallel  \cdot \vec{q}}$  \hspace{2cm}    & $\frac{q}{q_\perp }  \me^{2\mi d q_\perp} \left( R_s \frac{q_z^2}{q_\parallel^2}- R_p \frac{q_\perp^2 q_x^2}{q_\parallel^2 q^2} \right)\me^{\mi \delta \vec{r}_\parallel  \cdot \vec{q}}$.
\end{tabular}}

\subsection{Cavity setup: Including reflections from the crystal's front and back surfaces}

Similar to the last section, we can include the effect from reflections at the front and back surfaces of the crystal by using the appropriate scattering Green's tensor in addition to the bulk Green's tensor in Eq.~\eqref{eq:S2ofFilterAndG}. We first again neglect absorption by assuming that $n(\Omega)$ is real. The bulk contribution is not further restricted in this configuration and thus agrees with the signal considered in previous studies in which reflection effects have been neglected \cite{PRAlong2,lindel2020theory2}. It can be derived by using the bulk Green's tensor (Eq.~\eqref{eqapp:BulkGreensTensor}) in Eq.~\eqref{eq:S2ofFilterAndG} and the resulting expression is again given by Eq.~\eqref{eq:Signal2} with $q_\perp = q_z$, $O^{(j)}(\vec{q}, \delta \vec{r}_\parallel) = 1$, $N(d, \delta \vec{r}_\parallel) = N$ and \mbox{$p^{(0)}(\vec{q},\delta \vec{r}_\parallel) =q \frac{1-q^2_x/q^2}{2 q_\perp}\me^{\mi \vec{q} \cdot \delta \vec{r}_\parallel}$}. 
The scattering contribution is obtained by inserting the Green's tensor of the cavity in Eq.~\eqref{eqapp:GreensTensorCavity} into Eq.~\eqref{eq:S2ofFilterAndG}. After some calculation very similar to the ones in the last section one obtains 
\begin{multline} \label{eq:CavityG1}
g^{(1)}(\delta t, \delta \vec{r}_\parallel )  = C \int\limits_0^\infty\dif \Omega \, \mathrm{cos}(\Omega \delta t) E_\mathrm{vac}^{2}(\Omega)f^2(\Omega) \int \frac{\dif^2 q_{\parallel}}{4\pi q^2}  \me^{-(q_x^2+q_y^2)\beamwaist^2/4} \\\times \mathrm{Re}\left\{ \frac{q}{q_\perp} \me^{\mi \vec{q}\cdot \delta \vec{r}_\parallel} \me^{2\mi q_\perp L} \left(   \frac{R_s^2 q_y^2}{D_s q_\parallel^2} +   \frac{R_p^2 q_x^2 q_\perp^2}{D_p q^2q_\parallel^2} \right)\left(   \mathrm{sinc}^2\left[ \frac{L}{2} \left(n_g \frac{\Omega}{c}-q_z\right)\right] + \mathrm{sinc}^2\left[ \frac{L}{2} \left(n_g \frac{\Omega}{c}+q_z\right)\right]   \right)  \right. \\
\left. + \frac{2q}{q_\perp} \me^{\mi \vec{q}\cdot \delta \vec{r}_\parallel} \me^{\mi q_\perp L} \left(    \frac{R_s q_y^2}{D_s q_\parallel^2} -   \frac{R_p q_x^2 q_\perp^2}{D_p q^2q_\parallel^2} \right) \mathrm{sinc}\left[ \frac{L}{2} \left(n_g \frac{\Omega}{c}-q_z\right)\right] \mathrm{sinc}\left[ \frac{L}{2} \left(n_g \frac{\Omega}{c}+q_z\right)\right]    \right\}
\end{multline}
In the case that the crystal length is very short, e.g. set to $L = 1\,\mu$m (third row of Fig.~3c in the main plot), one needs to take all these terms into account. However, in case $L = 0.1\,$mm (as used throughout the rest of the paper) it is enough to only include the phase-matched contribution, i.e. only the term proportional to $ \mathrm{sinc}^2\left[L\left(n_g(\Omega/c)-q_z\right)/2\right] $. In this case we can again bring Eq.~\eqref{eq:CavityG1} into the form of Eq.~\eqref{eq:Signal2} with $q_\perp = q_z$, $O(\vec{q}, \delta \vec{r}_\parallel, d) = 1$, $N(d, \delta \vec{r}_\parallel) = N$, the response function given by Eq.~\eqref{eq:Response} and
\begin{align}
p^{(1)}(\vec{q},\delta \vec{r}_\parallel) =\frac{q}{q_\perp} \me^{\mi \vec{q}\cdot \delta \vec{r}_\parallel} \me^{2\mi q_\perp L} \left(   \frac{R_s^2 q_y^2}{D_s q_\parallel^2} +   \frac{R_p^2 q_x^2 q_\perp^2}{D_p q^2q_\parallel^2} \right).
\end{align}
Note, that this propagation factor only includes contributions which are at least quadratic in the reflection coefficients, meaning that the virtual photon is reflected at least twice. \\
In the case where the two pulses are assumed to propagate in opposite direction one has to replace $\me^{\mi k z}$ by $\me^{-\mi k z}$ in Eq.~\eqref{eq:E2}. The calculation is very similar to the one before. For the bulk contribution, the result is again given by Eq.~\eqref{eq:Signal2} with $q_\perp = q_z$, $O(\vec{q}, \delta \vec{r}_\parallel, d) = 1$, $N(d, \delta \vec{r}_\parallel) = N$ and $p^{(0)} =q \frac{1-q^2_x/q^2}{q_\perp}\me^{\mi \vec{q} \cdot \delta \vec{r}_\parallel}$ as for the case were the two pulses propagate into the same direction but the response function now reads
\begin{align}
R^2(\vec{q}) & =  \me^{-(q_x^2+q_y^2)\beamwaist^2/4}\mathrm{sinc}\left[ \frac{L}{2} \left(n_g \frac{\Omega}{c}-q_z\right)\right]\mathrm{sinc}\left[ \frac{L}{2} \left(n_g \frac{\Omega}{c}+q_z\right)\right] f(\Omega).
\end{align}
We see that there is no phase-matched contribution as expected since the two pulses propagate into opposite directions. 
The scattering contribution is also given by Eq.~\eqref{eq:Signal2} with $q_\perp = q_z$, $O(\vec{q}, \delta \vec{r}_\parallel, d) = 1$, $N(d, \delta \vec{r}_\parallel) = N$, the response function in Eq.~\eqref{eq:Response2} and
\begin{align}
p^{(1)} (\vec{q},\delta \vec{r}_\parallel)= \frac{q}{q_\perp} \me^{\mi \vec{q}\cdot \delta \vec{r}_\parallel} \me^{\mi q_\perp L} \left(    \frac{R_s q_y^2}{D_s q_\parallel^2} -   \frac{R_p q_x^2 q_\perp^2}{D_p q^2q_\parallel^2} \right) .
\end{align}
Note, that this propagation factor now includes contribution which are proportional to $R_\sigma$, meaning that the virtual photon can this time also be reflected only once and still be phase-matched as expected. \\

Lastly, we calculate the signal accounting not only for reflections at the front and back surfaces of the crystal but also for absorption in the THz region. To do so we allow $n(\Omega)$ to have a nonvanishing imaginary part (compare right hand side of Fig.~\ref{fig:Refr}). For the bulk contribution with both pulses propagating into the same direction one finds as previously also obtained in Ref.~\cite{lindel2020theory2}
\begin{multline}
g^{(0)}(\delta t, \delta \vec{r}_\parallel )  = C \int\limits_0^\infty\dif \Omega \, \mathrm{cos}(\Omega \delta t) E_\mathrm{vac}^{2}(\Omega) f^2(\Omega) \int \frac{\dif^2 q_{\parallel}}{4\pi \mathrm{Re}[q]}  \me^{-(q_x^2 + q_y^2)\beamwaist^2/4}\\
\times \mathrm{Re}\left[\me^{\mi \vec{q} \cdot \delta \vec{r}_\parallel} \frac{\left(1-\frac{q_x^2}{q^2}\right) }{q_z L^2}\left( \frac{\mi L}{q_z -n_g\Omega/c} +  \frac{1- \me^{\mi L(q_z -n_g\Omega/c)}}{  (q_z -n_g\Omega/c)^2}    \right) \right].
 \end{multline}
 Here, only the phase-matched contribution is included.  \\
 For the scattering contribution we repeat the same calculation leading to Eq.~\eqref{eq:CavityG1} but allowing for a imaginary part of $n(\Omega)$. This leads to 
\begin{multline} \label{eq:signalCavityG1}
g^{(1)}(\delta t, \delta \vec{r}_\parallel )  =C \int\limits_0^\infty\dif \Omega \, \mathrm{cos}(\Omega \delta t) E_\mathrm{vac}^{2}(\Omega) f^2(\Omega) \int \frac{\dif^2 q_{\parallel}}{4\pi \mathrm{Re}[q]}\me^{-(q_x^2 + q_y^2)\beamwaist^2/4}\\
\times \mathrm{Re}\left\{ \frac{1}{q_\perp} \me^{\mi \vec{q}\cdot \delta \vec{r}_\parallel} \me^{2\mi q_\perp L} \left(   \frac{R_s^2 q_y^2}{D_s q_\parallel^2} +   \frac{R_p^2 q_x^2 q_\perp^2}{D_p q^2q_\parallel^2} \right)\left(   \mathrm{sinc}^2\left[ \frac{L}{2} \left(n_g \frac{\Omega}{c}-q_z\right)\right] + \mathrm{sinc}^2\left[ \frac{L}{2} \left(n_g \frac{\Omega}{c}+q_z\right)\right]   \right)  \right. \\
\left. + \frac{2}{q_\perp} \me^{\mi \vec{q}\cdot \delta \vec{r}_\parallel} \me^{\mi q_\perp L} \left(    \frac{R_s q_y^2}{D_s q_\parallel^2} -   \frac{R_p q_x^2 q_\perp^2}{D_p q^2q_\parallel^2} \right) \mathrm{sinc}\left[ \frac{L}{2} \left(n_g \frac{\Omega}{c}-q_z\right)\right] \mathrm{sinc}\left[ \frac{L}{2} \left(n_g \frac{\Omega}{c}+q_z\right)\right]    \right\}.
\end{multline}
In the case where the two pulses propagate into opposite direction and absorption effects are considered, the bulk contribution to the signal is given by
\begin{multline}
g^{(0)}(\delta t, \delta \vec{r}_\parallel )  = C \int\limits_0^\infty\dif \Omega \, \mathrm{cos}(\Omega \delta t) E_\mathrm{vac}^{2}(\Omega) f^2(\Omega)\int \frac{\dif^2 q_{\parallel}}{4\pi \mathrm{Re}[q]}  \me^{-(q_x^2 + q_y^2)\beamwaist^2/4}\\
\times \mathrm{Re}\left[\me^{\mi \vec{q} \cdot \delta \vec{r}_\parallel} \frac{\left(1-\frac{q_x^2}{q^2}\right) }{q_z L^2}\left( \frac{\mi\frac{q_z c}{n_g \Omega}\mathrm{sin}(Ln_g \Omega/c) + \mathrm{cos}(Ln_g \Omega/c) - \me^{\mi L q_z}}{(q_z -n_g\Omega/c)(q_z + n_g \Omega/c)}   \right) \right].
 \end{multline}
and the scattering contribution by
\begin{multline} \label{eq:signalCavityOpp}
g^{(1)}(\delta t, \delta \vec{r}_\parallel )  =C \int\limits_0^\infty\dif \Omega \, \mathrm{cos}(\Omega \delta t) E_\mathrm{vac}^{2}(\Omega)f^2(\Omega) \int \frac{\dif^2 q_{\parallel}}{4\pi \mathrm{Re}[q]}\me^{-(q_x^2 + q_y^2)\beamwaist^2/4}\\
\times \mathrm{Re}\left\{ \frac{2}{q_\perp} \me^{\mi \vec{q}\cdot \delta \vec{r}_\parallel} \me^{2\mi q_\perp L} \left(   \frac{R_s^2 q_y^2}{D_s q_\parallel^2} +   \frac{R_p^2 q_x^2 q_\perp^2}{D_p q^2q_\parallel^2} \right) \mathrm{sinc}\left[ \frac{L}{2} \left(n_g \frac{\Omega}{c}-q_z\right)\right] \mathrm{sinc}\left[ \frac{L}{2} \left(n_g \frac{\Omega}{c}+q_z\right)\right]   \right. \\
\left. +  \frac{1}{q_\perp} \me^{\mi \vec{q}\cdot \delta \vec{r}_\parallel} \me^{\mi q_\perp L} \left(    \frac{R_s q_y^2}{D_s q_\parallel^2} -   \frac{R_p q_x^2 q_\perp^2}{D_p q^2q_\parallel^2} \right)   \left(   \mathrm{sinc}^2\left[ \frac{L}{2} \left(n_g \frac{\Omega}{c}-q_z\right)\right] + \mathrm{sinc}^2\left[ \frac{L}{2} \left(n_g \frac{\Omega}{c}+q_z\right)\right]   \right)       \right\}.
\end{multline}

\section{Microscopic Interpretation }\label{app:MicroscopicInterp}

We discuss the microscopic processes leading to the contribution to the signal's variance found in Eq.~\eqref{eq:S2ofFilter2}. We first realize that, on the one hand, terms of the structure $\hat{E}_x(\vec{r}^\prime,\Omega)\hat{E}_y(\vec{r}^\prime,\omega-\Omega)G_{xx}(\vec{r},\vec{r}^\prime,\omega)$ correspond to a processes generally referred to as sum-frequency generation \cite{Boyd20032}. Here an atom absorbs two photons with polarization $x,y$ and frequency $\Omega,\omega-\Omega $ respectively and the excited atom subsequently emits an $x$ polarized photon with frequency $\omega$. On the other hand, terms of the structure $\hat{E}^\dagger_x(\vec{r}^\prime,\Omega)\hat{E}_y(\vec{r}^\prime,\omega+\Omega)G_{xx}(\vec{r},\vec{r}^\prime,\omega)$ describe the process of parametric down-conversion \cite{Boyd20032}, where an atom is excited by one photon of frequency $\omega+ \Omega$ which subsequently deexcites by a two-photon emission process where both photons are $x$-polarised and have frequencies $\Omega$ and $\omega$, respectively. \\
Now let us turn our attention to Eq.~\eqref{eq:S2g}, i.e. 
\begin{multline} \label{eq:S2gg}
g = \mathcal{S}\left\{ \left( 4\pi \epsilon_0 c \int_0^\infty \dif \omega \frac{\eta(\omega)}{\hbar \omega} \int \dif^2 r_\parallel  \left[ \mi \hat{E}_{y,1}^\dagger(\vec{r}_\parallel, \omega) \hat{E}_{x,1}(\vec{r}_\parallel, \omega) + \text{h.c.} \right] \right)\right.  \\\left.
\times \left( 4\pi \epsilon_0 c \int_0^\infty \dif \omega \frac{\eta(\omega)}{\hbar \omega} \int \dif^2 r_\parallel  \left[ \mi \hat{E}_{y,2}^\dagger(\vec{r}_\parallel, \omega) \hat{E}_{x,2}(\vec{r}_\parallel, \omega) + \text{h.c.} \right] \right) \right\}.
\end{multline}
Note, that in the lowest order in $\chi^{(2)}$, $\hat{E}_x(\vec{r}, \omega)$ is given by \cite{PRAlong2}
\begin{align} \label{eq:Ex1}
\hat{E}_x(\vec{r}, \omega) = \chi^{(2)} \omega^2 \mu_0 \int_{V_\textrm{C}} \dif^3 r^\prime \, G_{xx}(\vec{r},\vec{r}^\prime,\omega)  \int_{-\infty}^\infty \dif \Omega \,\,    \hat{E}_{\textrm{vac},x}(\vec{r}^\prime,\Omega)  \hat{E}_{\textrm{p},y}(\vec{r}^\prime,\omega -\Omega )
\end{align}
 such that it is apparent from the previous discussion that this field stems from the nonlinear process of either parametric down-conversion or sum-frequency generation depending on the sign of the frequencies. When taking the ground state expectation value only terms proportional to $\hat{E}_{\textrm{vac},x}(\Omega)\hat{E}^\dagger_{\textrm{vac},x}(\Omega)$ contribute where $\Omega$ is now positive. This means that on a microscopic level only those nonlinear processes contribute where a (virtual) photon at frequency $\Omega$ is generated in a first nonlinear process and subsequently absorbed by a second one. To illustrate this, we insert Eq.~\eqref{eq:Ex1} into Eq.~\eqref{eq:S2g} and consider one of the resulting terms given by
\begin{align}
\hat{E}_x(\vec{r}^{\prime\prime},\Omega)\hat{E}_y(\vec{r}^{\prime\prime},\Omega-\omega^\prime)G_{xx}(\vec{r},\vec{r}^{\prime\prime},\omega^\prime)\hat{E}^\dagger_x(\vec{r}^\prime,\Omega)\hat{E}_y(\vec{r}^\prime,\Omega+\omega)G_{xx}(\vec{r},\vec{r}^\prime,\omega).
\end{align}
It is apparent that this describes the processes of spontaneous parametric down-conversion at position $\vec{r}^\prime$ and sum-frequency generation of one of the generated photon together with a photon of the laser pulse at position $\vec{r}^{\prime\prime}$. \\

\end{widetext}


\begin{thebibliography}{20}%
\makeatletter
\providecommand \@ifxundefined [1]{%
 \@ifx{#1\undefined}
}%
\providecommand \@ifnum [1]{%
 \ifnum #1\expandafter \@firstoftwo
 \else \expandafter \@secondoftwo
 \fi
}%
\providecommand \@ifx [1]{%
 \ifx #1\expandafter \@firstoftwo
 \else \expandafter \@secondoftwo
 \fi
}%
\providecommand \natexlab [1]{#1}%
\providecommand \enquote  [1]{``#1''}%
\providecommand \bibnamefont  [1]{#1}%
\providecommand \bibfnamefont [1]{#1}%
\providecommand \citenamefont [1]{#1}%
\providecommand \href@noop [0]{\@secondoftwo}%
\providecommand \href [0]{\begingroup \@sanitize@url \@href}%
\providecommand \@href[1]{\@@startlink{#1}\@@href}%
\providecommand \@@href[1]{\endgroup#1\@@endlink}%
\providecommand \@sanitize@url [0]{\catcode `\\12\catcode `\$12\catcode
  `\&12\catcode `\#12\catcode `\^12\catcode `\_12\catcode `\%12\relax}%
\providecommand \@@startlink[1]{}%
\providecommand \@@endlink[0]{}%
\providecommand \url  [0]{\begingroup\@sanitize@url \@url }%
\providecommand \@url [1]{\endgroup\@href {#1}{\urlprefix }}%
\providecommand \urlprefix  [0]{URL }%
\providecommand \Eprint [0]{\href }%
\providecommand \doibase [0]{http://dx.doi.org/}%
\providecommand \selectlanguage [0]{\@gobble}%
\providecommand \bibinfo  [0]{\@secondoftwo}%
\providecommand \bibfield  [0]{\@secondoftwo}%
\providecommand \translation [1]{[#1]}%
\providecommand \BibitemOpen [0]{}%
\providecommand \bibitemStop [0]{}%
\providecommand \bibitemNoStop [0]{.\EOS\space}%
\providecommand \EOS [0]{\spacefactor3000\relax}%
\providecommand \BibitemShut  [1]{\csname bibitem#1\endcsname}%
\let\auto@bib@innerbib\@empty
\bibitem [{\citenamefont {Lamb}\ and\ \citenamefont
  {Retherford}(1947)}]{Lamb1947}%
  \BibitemOpen
  \bibfield  {author} {\bibinfo {author} {\bibfnamefont {W.~E.}\ \bibnamefont
  {Lamb}}\ and\ \bibinfo {author} {\bibfnamefont {R.~C.}\ \bibnamefont
  {Retherford}},\ }\href {\doibase 10.1103/PhysRev.72.241} {\bibfield
  {journal} {\bibinfo  {journal} {Phys. Rev.}\ }\textbf {\bibinfo {volume}
  {72}},\ \bibinfo {pages} {241} (\bibinfo {year} {1947})}\BibitemShut
  {NoStop}%
\bibitem [{\citenamefont {Casimir}(1948)}]{Casimir1948}%
  \BibitemOpen
  \bibfield  {author} {\bibinfo {author} {\bibfnamefont {H.~B.~G.}\
  \bibnamefont {Casimir}},\ }\href {\doibase citeulike-article-id:8810715}
  {\bibfield  {journal} {\bibinfo  {journal} {Proc. K. Ned. Akad.}\ }\textbf
  {\bibinfo {volume} {360}},\ \bibinfo {pages} {793} (\bibinfo {year}
  {1948})}\BibitemShut {NoStop}%
\bibitem [{\citenamefont {Milonni}(1994)}]{Milonni1994}%
  \BibitemOpen
  \bibfield  {author} {\bibinfo {author} {\bibfnamefont {P.~W.}\ \bibnamefont
  {Milonni}},\ }\href
  {https://books.google.ru/books?id=uPHJCgAAQBAJ&printsec=frontcover&dq=milonni+quantum+vacuum&hl=en&sa=X&ved=0ahUKEwiD3PSQsojUAhXKDpoKHbpwD7wQ6AEIJjAA#v=onepage&q=milonni
  quantum vacuum&f=false} {\emph {\bibinfo {title} {{The quantum vacuum: an
  introduction to quantum electrodynamics}}}}\ (\bibinfo  {publisher} {Academic
  Press, San Diego},\ \bibinfo {year} {1994})\BibitemShut {NoStop}%
\bibitem [{\citenamefont {Purcell}(1946)}]{Purcell1946}%
  \BibitemOpen
  \bibfield  {author} {\bibinfo {author} {\bibfnamefont {E.~M.}\ \bibnamefont
  {Purcell}},\ }\href {\doibase 10.1103/PhysRev.69.674.2} {\bibfield  {journal}
  {\bibinfo  {journal} {Proc. Am. Phys. Soc.}\ }\textbf {\bibinfo {volume}
  {69}},\ \bibinfo {pages} {674} (\bibinfo {year} {1946})}\BibitemShut
  {NoStop}%
\bibitem [{\citenamefont {Hemmerich}\ \emph {et~al.}(2018)\citenamefont
  {Hemmerich}, \citenamefont {Bennett},\ and\ \citenamefont
  {Buhmann}}]{Hemmerich2018}%
  \BibitemOpen
  \bibfield  {author} {\bibinfo {author} {\bibfnamefont {J.~L.}\ \bibnamefont
  {Hemmerich}}, \bibinfo {author} {\bibfnamefont {R.}~\bibnamefont {Bennett}},
  \ and\ \bibinfo {author} {\bibfnamefont {S.~Y.}\ \bibnamefont {Buhmann}},\
  }\href {\doibase 10.1038/s41467-018-05091-x} {\bibfield  {journal} {\bibinfo
  {journal} {Nat. Commun.}\ }\textbf {\bibinfo {volume} {9}},\ \bibinfo {pages}
  {2934} (\bibinfo {year} {2018})}\BibitemShut {NoStop}%
\bibitem [{\citenamefont {Hutchison}\ \emph {et~al.}(2012)\citenamefont
  {Hutchison}, \citenamefont {Schwartz}, \citenamefont {Genet}, \citenamefont
  {Devaux},\ and\ \citenamefont {Ebbesen}}]{hutchison2012modifying}%
  \BibitemOpen
  \bibfield  {author} {\bibinfo {author} {\bibfnamefont {J.~A.}\ \bibnamefont
  {Hutchison}}, \bibinfo {author} {\bibfnamefont {T.}~\bibnamefont {Schwartz}},
  \bibinfo {author} {\bibfnamefont {C.}~\bibnamefont {Genet}}, \bibinfo
  {author} {\bibfnamefont {E.}~\bibnamefont {Devaux}}, \ and\ \bibinfo {author}
  {\bibfnamefont {T.~W.}\ \bibnamefont {Ebbesen}},\ }\href
  {https://onlinelibrary.wiley.com/doi/full/10.1002/anie.201107033} {\bibfield
  {journal} {\bibinfo  {journal} {Angew. Chem., Int. Ed.}\ }\textbf {\bibinfo
  {volume} {51}},\ \bibinfo {pages} {1592} (\bibinfo {year}
  {2012})}\BibitemShut {NoStop}%
\bibitem [{\citenamefont {Ribeiro}\ \emph {et~al.}(2018)\citenamefont
  {Ribeiro}, \citenamefont {Mart{\'\i}nez-Mart{\'\i}nez}, \citenamefont {Du},
  \citenamefont {Campos-Gonzalez-Angulo},\ and\ \citenamefont
  {Yuen-Zhou}}]{ribeiro2018polariton}%
  \BibitemOpen
  \bibfield  {author} {\bibinfo {author} {\bibfnamefont {R.~F.}\ \bibnamefont
  {Ribeiro}}, \bibinfo {author} {\bibfnamefont {L.~A.}\ \bibnamefont
  {Mart{\'\i}nez-Mart{\'\i}nez}}, \bibinfo {author} {\bibfnamefont
  {M.}~\bibnamefont {Du}}, \bibinfo {author} {\bibfnamefont {J.}~\bibnamefont
  {Campos-Gonzalez-Angulo}}, \ and\ \bibinfo {author} {\bibfnamefont
  {J.}~\bibnamefont {Yuen-Zhou}},\ }\href
  {https://pubs.rsc.org/en/content/articlelanding/sc/2018/c8sc01043a#!divAbstract}
  {\bibfield  {journal} {\bibinfo  {journal} {Chem. Sci.}\ }\textbf {\bibinfo
  {volume} {9}},\ \bibinfo {pages} {6325} (\bibinfo {year} {2018})}\BibitemShut
  {NoStop}%
\bibitem [{\citenamefont {Riek}\ \emph {et~al.}(2015)\citenamefont {Riek},
  \citenamefont {Seletskiy}, \citenamefont {Moskalenko}, \citenamefont
  {Schmidt}, \citenamefont {Krauspe}, \citenamefont {Eckart}, \citenamefont
  {Eggert}, \citenamefont {Burkard},\ and\ \citenamefont
  {Leitenstorfer}}]{Riek2015}%
  \BibitemOpen
  \bibfield  {author} {\bibinfo {author} {\bibfnamefont {C.}~\bibnamefont
  {Riek}}, \bibinfo {author} {\bibfnamefont {D.~V.}\ \bibnamefont {Seletskiy}},
  \bibinfo {author} {\bibfnamefont {A.~S.}\ \bibnamefont {Moskalenko}},
  \bibinfo {author} {\bibfnamefont {J.~F.}\ \bibnamefont {Schmidt}}, \bibinfo
  {author} {\bibfnamefont {P.}~\bibnamefont {Krauspe}}, \bibinfo {author}
  {\bibfnamefont {S.}~\bibnamefont {Eckart}}, \bibinfo {author} {\bibfnamefont
  {S.}~\bibnamefont {Eggert}}, \bibinfo {author} {\bibfnamefont
  {G.}~\bibnamefont {Burkard}}, \ and\ \bibinfo {author} {\bibfnamefont
  {A.}~\bibnamefont {Leitenstorfer}},\ }\href {\doibase
  10.1126/science.aac9788} {\bibfield  {journal} {\bibinfo  {journal}
  {Science}\ }\textbf {\bibinfo {volume} {350}},\ \bibinfo {pages} {420}
  (\bibinfo {year} {2015})}\BibitemShut {NoStop}%
\bibitem [{\citenamefont {Benea-Chelmus}\ \emph {et~al.}(2019)\citenamefont
  {Benea-Chelmus}, \citenamefont {Settembrini}, \citenamefont {Scalari},\ and\
  \citenamefont {Faist}}]{Benea-Chelmus2019}%
  \BibitemOpen
  \bibfield  {author} {\bibinfo {author} {\bibfnamefont {I.-C.}\ \bibnamefont
  {Benea-Chelmus}}, \bibinfo {author} {\bibfnamefont {F.~F.}\ \bibnamefont
  {Settembrini}}, \bibinfo {author} {\bibfnamefont {G.}~\bibnamefont
  {Scalari}}, \ and\ \bibinfo {author} {\bibfnamefont {J.}~\bibnamefont
  {Faist}},\ }\href {\doibase 10.1038/s41586-019-1083-9} {\bibfield  {journal}
  {\bibinfo  {journal} {Nature}\ }\textbf {\bibinfo {volume} {568}},\ \bibinfo
  {pages} {202} (\bibinfo {year} {2019})}\BibitemShut {NoStop}%
\bibitem []{Seletzkiy}%
  \BibitemOpen
  \bibfield  {author} {\bibinfo {author} {\bibfnamefont {S.}\ \bibnamefont
  {Virally}}, \bibinfo {author} {\bibfnamefont {P.}\ \bibnamefont
  {Cusson}},\ and\ \bibinfo {author} {\bibfnamefont {D. V.}~\bibnamefont
  {Seletskiy}},\ }\href {https://arxiv.org/abs/2106.04402} {\bibfield  {journal}
  {\bibinfo  {journal} {arXiv preprint arXiv:2106.04402}\ }(\bibinfo {year} {2021})}\BibitemShut {NoStop}%
\bibitem [{\citenamefont {Moskalenko}\ \emph {et~al.}(2015)\citenamefont
  {Moskalenko}, \citenamefont {Riek}, \citenamefont {Seletskiy}, \citenamefont
  {Burkard},\ and\ \citenamefont {Leitenstorfer}}]{Moskalenko2015}%
  \BibitemOpen
  \bibfield  {author} {\bibinfo {author} {\bibfnamefont {A.~S.}\ \bibnamefont
  {Moskalenko}}, \bibinfo {author} {\bibfnamefont {C.}~\bibnamefont {Riek}},
  \bibinfo {author} {\bibfnamefont {D.~V.}\ \bibnamefont {Seletskiy}}, \bibinfo
  {author} {\bibfnamefont {G.}~\bibnamefont {Burkard}}, \ and\ \bibinfo
  {author} {\bibfnamefont {A.}~\bibnamefont {Leitenstorfer}},\ }\href {\doibase
  10.1103/PhysRevLett.115.263601} {\bibfield  {journal} {\bibinfo  {journal}
  {Phys. Rev. Lett.}\ }\textbf {\bibinfo {volume} {115}},\ \bibinfo {pages}
  {263601} (\bibinfo {year} {2015})}\BibitemShut {NoStop}%
\bibitem [{\citenamefont {Riek}\ \emph {et~al.}(2017)\citenamefont {Riek},
  \citenamefont {Sulzer}, \citenamefont {Seeger}, \citenamefont {Moskalenko},
  \citenamefont {Burkard}, \citenamefont {Seletskiy},\ and\ \citenamefont
  {Leitenstorfer}}]{riek2017subcycle}%
  \BibitemOpen
  \bibfield  {author} {\bibinfo {author} {\bibfnamefont {C.}~\bibnamefont
  {Riek}}, \bibinfo {author} {\bibfnamefont {P.}~\bibnamefont {Sulzer}},
  \bibinfo {author} {\bibfnamefont {M.}~\bibnamefont {Seeger}}, \bibinfo
  {author} {\bibfnamefont {A.~S.}\ \bibnamefont {Moskalenko}}, \bibinfo
  {author} {\bibfnamefont {G.}~\bibnamefont {Burkard}}, \bibinfo {author}
  {\bibfnamefont {D.~V.}\ \bibnamefont {Seletskiy}}, \ and\ \bibinfo {author}
  {\bibfnamefont {A.}~\bibnamefont {Leitenstorfer}},\ }\href
  {https://www.nature.com/articles/nature21024} {\bibfield  {journal} {\bibinfo
   {journal} {Nature}\ }\textbf {\bibinfo {volume} {541}},\ \bibinfo {pages}
  {376} (\bibinfo {year} {2017})}\BibitemShut {NoStop}%
\bibitem [{\citenamefont {Kizmann}\ \emph {et~al.}(2019)\citenamefont
  {Kizmann}, \citenamefont {Guedes}, \citenamefont {Seletskiy}, \citenamefont
  {Moskalenko}, \citenamefont {Leitenstorfer},\ and\ \citenamefont
  {Burkard}}]{kizmann2019subcycle}%
  \BibitemOpen
  \bibfield  {author} {\bibinfo {author} {\bibfnamefont {M.}~\bibnamefont
  {Kizmann}}, \bibinfo {author} {\bibfnamefont {T.~L. d.~M.}\ \bibnamefont
  {Guedes}}, \bibinfo {author} {\bibfnamefont {D.~V.}\ \bibnamefont
  {Seletskiy}}, \bibinfo {author} {\bibfnamefont {A.~S.}\ \bibnamefont
  {Moskalenko}}, \bibinfo {author} {\bibfnamefont {A.}~\bibnamefont
  {Leitenstorfer}}, \ and\ \bibinfo {author} {\bibfnamefont {G.}~\bibnamefont
  {Burkard}},\ }\href {https://www.nature.com/articles/s41567-019-0560-2}
  {\bibfield  {journal} {\bibinfo  {journal} {Nature Physics}\ }\textbf
  {\bibinfo {volume} {15}},\ \bibinfo {pages} {960} (\bibinfo {year}
  {2019})}\BibitemShut {NoStop}%
\bibitem [{\citenamefont {Lindel}\ \emph {et~al.}(2020)\citenamefont {Lindel},
  \citenamefont {Bennett},\ and\ \citenamefont {Buhmann}}]{lindel2020theory}%
  \BibitemOpen
  \bibfield  {author} {\bibinfo {author} {\bibfnamefont {F.}~\bibnamefont
  {Lindel}}, \bibinfo {author} {\bibfnamefont {R.}~\bibnamefont {Bennett}}, \
  and\ \bibinfo {author} {\bibfnamefont {S.~Y.}\ \bibnamefont {Buhmann}},\
  }\href {https://journals.aps.org/pra/abstract/10.1103/PhysRevA.102.041701}
  {\bibfield  {journal} {\bibinfo  {journal} {Physical Review A}\ }\textbf
  {\bibinfo {volume} {102}},\ \bibinfo {pages} {041701} (\bibinfo {year}
  {2020})}\BibitemShut {NoStop}%
\bibitem [{\citenamefont {Lindel}\ \emph
  {et~al.}(2021{\natexlab{a}})\citenamefont {Lindel}, \citenamefont {Bennett},\
  and\ \citenamefont {Buhmann}}]{PRAlong}%
  \BibitemOpen
  \bibfield  {author} {\bibinfo {author} {\bibfnamefont {F.}~\bibnamefont
  {Lindel}}, \bibinfo {author} {\bibfnamefont {R.}~\bibnamefont {Bennett}}, \
  and\ \bibinfo {author} {\bibfnamefont {S.~Y.}\ \bibnamefont {Buhmann}},\
  }\href {https://journals.aps.org/pra/abstract/10.1103/PhysRevA.103.033705}
  {\bibfield  {journal} {\bibinfo  {journal} {Phys. Rev. A}\ }\textbf {\bibinfo
  {volume} {103}},\ \bibinfo {pages} {033705} (\bibinfo {year}
  {2021}{\natexlab{a}})}\BibitemShut {NoStop}%
\bibitem [{\citenamefont {Lindel}\ \emph
  {et~al.}(2021{\natexlab{b}})\citenamefont {Lindel}, \citenamefont {Bennett},\
  and\ \citenamefont {Buhmann}}]{SUPP}%
  \BibitemOpen
  \bibfield  {author} {\bibinfo {author} {\bibfnamefont {See Supplemental Material for derivations and further discussions.}}}
\bibitem [{\citenamefont {Buhmann}(2012)}]{Buhmann2012BothBooks}%
  \BibitemOpen
  \bibfield  {author} {\bibinfo {author} {\bibfnamefont {S.~Y.}\ \bibnamefont
  {Buhmann}},\ }\href@noop {} {\emph {\bibinfo {title} {{Dispersion
  Forces}}}},\ \bibinfo {series} {Springer Tracts in Modern Physics}, Vol.\
  \bibinfo {volume} {247}\ (\bibinfo  {publisher} {Springer},\ \bibinfo
  {address} {Berlin, Heidelberg},\ \bibinfo {year} {2012})\BibitemShut
  {NoStop}%
\bibitem [{\citenamefont {Leitenstorfer}\ \emph {et~al.}(1999)\citenamefont
  {Leitenstorfer}, \citenamefont {Hunsche}, \citenamefont {Shah}, \citenamefont
  {Nuss},\ and\ \citenamefont {Knox}}]{Leitenstorfer1999}%
  \BibitemOpen
  \bibfield  {author} {\bibinfo {author} {\bibfnamefont {A.}~\bibnamefont
  {Leitenstorfer}}, \bibinfo {author} {\bibfnamefont {S.}~\bibnamefont
  {Hunsche}}, \bibinfo {author} {\bibfnamefont {J.}~\bibnamefont {Shah}},
  \bibinfo {author} {\bibfnamefont {M.~C.}\ \bibnamefont {Nuss}}, \ and\
  \bibinfo {author} {\bibfnamefont {W.~H.}\ \bibnamefont {Knox}},\ }\href
  {\doibase 10.1063/1.123601} {\bibfield  {journal} {\bibinfo  {journal} {Appl.
  Phys. Lett.}\ }\textbf {\bibinfo {volume} {74}},\ \bibinfo {pages} {1516}
  (\bibinfo {year} {1999})}\BibitemShut {NoStop}%
\bibitem [{\citenamefont {Rizzuto}\ \emph {et~al.}(2004)\citenamefont
  {Rizzuto}, \citenamefont {Passante},\ and\ \citenamefont
  {Persico}}]{rizzuto2004dynamical}%
  \BibitemOpen
  \bibfield  {author} {\bibinfo {author} {\bibfnamefont {L.}~\bibnamefont
  {Rizzuto}}, \bibinfo {author} {\bibfnamefont {R.}~\bibnamefont {Passante}}, \
  and\ \bibinfo {author} {\bibfnamefont {F.}~\bibnamefont {Persico}},\
  }\href@noop {} {\bibfield  {journal} {\bibinfo  {journal} {Physical Review
  A}\ }\textbf {\bibinfo {volume} {70}},\ \bibinfo {pages} {012107} (\bibinfo
  {year} {2004})}\BibitemShut {NoStop}%
\bibitem [{\citenamefont {Passante}\ \emph {et~al.}(2006)\citenamefont
  {Passante}, \citenamefont {Persico},\ and\ \citenamefont
  {Rizzuto}}]{passante2006causality}%
  \BibitemOpen
  \bibfield  {author} {\bibinfo {author} {\bibfnamefont {R.}~\bibnamefont
  {Passante}}, \bibinfo {author} {\bibfnamefont {F.}~\bibnamefont {Persico}}, \
  and\ \bibinfo {author} {\bibfnamefont {L.}~\bibnamefont {Rizzuto}},\ }\href
  {https://iopscience.iop.org/article/10.1088/0953-4075/39/15/S15/meta}
  {\bibfield  {journal} {\bibinfo  {journal} {J. Phys. B}\ }\textbf {\bibinfo
  {volume} {39}},\ \bibinfo {pages} {S685} (\bibinfo {year}
  {2006})}\BibitemShut {NoStop}%
\bibitem [{\citenamefont {Benea-Chelmus}\ \emph {et~al.}(2020)\citenamefont
  {Benea-Chelmus}, \citenamefont {Salamin}, \citenamefont {Settembrini},
  \citenamefont {Fedoryshyn}, \citenamefont {Heni}, \citenamefont {Elder},
  \citenamefont {Dalton}, \citenamefont {Leuthold},\ and\ \citenamefont
  {Faist}}]{Biswas}%
  \BibitemOpen
  \bibfield  {author} {\bibinfo {author} {\bibfnamefont {A. K.}\ \bibnamefont
  {Biswas}}, \bibinfo {author} {\bibfnamefont {G.}~\bibnamefont
  {Compagno}}, \bibinfo {author} {\bibfnamefont {G. M.}\ \bibnamefont
  {Palma}}, \bibinfo {author} {\bibfnamefont {R.}~\bibnamefont
  {Passante}}, \ and\ \bibinfo {author}
  {\bibfnamefont {F.}~\bibnamefont {Persico}},\ }\href
  {https://journals.aps.org/pra/abstract/10.1103/PhysRevA.42.4291}
  {\bibfield  {journal} {\bibinfo  {journal} {Phys. Rev. A}\ }\textbf {\bibinfo
  {volume} {42}},\ \bibinfo {pages} {4291} (\bibinfo {year} {1990})}\BibitemShut
  {NoStop}%
\bibitem [{\citenamefont {A.}\ \emph {et~al.}(2020)\citenamefont
  {Benea-Chelmus}, \citenamefont {Salamin}, \citenamefont {Settembrini},
  \citenamefont {Fedoryshyn}, \citenamefont {Heni}, \citenamefont {Elder},
  \citenamefont {Dalton}, \citenamefont {Leuthold},\ and\ \citenamefont
  {Faist}}]{Valentini}%
  \BibitemOpen
  \bibfield  {author} {\bibinfo {author} {\bibfnamefont {A.}\ \bibnamefont
  {Valentini}}}, \href
  {https://www.sciencedirect.com/science/article/pii/0375960191909525}
  {\bibfield  {journal} {\bibinfo  {journal} {Phys. Rev. A}\ }\textbf {\bibinfo
  {volume} {153}},\ \bibinfo {pages} {321325} (\bibinfo {year} {1991})}\BibitemShut
  {NoStop}%
\bibitem [{\citenamefont {Benea-Chelmus}\ \emph {et~al.}(2020)\citenamefont
  {Benea-Chelmus}, \citenamefont {Salamin}, \citenamefont {Settembrini},
  \citenamefont {Fedoryshyn}, \citenamefont {Heni}, \citenamefont {Elder},
  \citenamefont {Dalton}, \citenamefont {Leuthold},\ and\ \citenamefont
  {Faist}}]{Reznik1}%
  \BibitemOpen
  \bibfield  {author} {\bibinfo {author} {\bibfnamefont {B.}\ \bibnamefont
  {Reznik}}}, \href
  {https://link.springer.com/article/10.1023/A:1022875910744}
  {\bibfield  {journal} {\bibinfo  {journal} {Found. Phys.}\ }\textbf {\bibinfo
  {volume} {33}},\ \bibinfo {pages} {167} (\bibinfo {year} {2003})}\BibitemShut
  {NoStop}%
\bibitem [{\citenamefont {Benea-Chelmus}\ \emph {et~al.}(2020)\citenamefont
  {Benea-Chelmus}, \citenamefont {Salamin}, \citenamefont {Settembrini},
  \citenamefont {Fedoryshyn}, \citenamefont {Heni}, \citenamefont {Elder},
  \citenamefont {Dalton}, \citenamefont {Leuthold},\ and\ \citenamefont
  {Faist}}]{Martinez}%
  \BibitemOpen
  \bibfield  {author} {\bibinfo {author} {\bibfnamefont {A.}\ \bibnamefont
  {Pozas-Kerstjens}}, \ and\ \bibinfo {author}
  {\bibfnamefont {E.}~\bibnamefont {Mart{\'\i}n-Mart{\'\i}nez}},\ }\href
  {https://journals.aps.org/prd/abstract/10.1103/PhysRevD.92.064042}
  {\bibfield  {journal} {\bibinfo  {journal} {Phys. Rev. D}\ }\textbf {\bibinfo
  {volume} {92}},\ \bibinfo {pages} {064042} (\bibinfo {year} {2015})}\BibitemShut
  {NoStop}%
\bibitem [{\citenamefont {Benea-Chelmus}\ \emph {et~al.}(2020)\citenamefont
  {Benea-Chelmus}, \citenamefont {Salamin}, \citenamefont {Settembrini},
  \citenamefont {Fedoryshyn}, \citenamefont {Heni}, \citenamefont {Elder},
  \citenamefont {Dalton}, \citenamefont {Leuthold},\ and\ \citenamefont
  {Faist}}]{benea2020electro}%
  \BibitemOpen
  \bibfield  {author} {\bibinfo {author} {\bibfnamefont {I.-C.}\ \bibnamefont
  {Benea-Chelmus}}, \bibinfo {author} {\bibfnamefont {Y.}~\bibnamefont
  {Salamin}}, \bibinfo {author} {\bibfnamefont {F.~F.}\ \bibnamefont
  {Settembrini}}, \bibinfo {author} {\bibfnamefont {Y.}~\bibnamefont
  {Fedoryshyn}}, \bibinfo {author} {\bibfnamefont {W.}~\bibnamefont {Heni}},
  \bibinfo {author} {\bibfnamefont {D.~L.}\ \bibnamefont {Elder}}, \bibinfo
  {author} {\bibfnamefont {L.~R.}\ \bibnamefont {Dalton}}, \bibinfo {author}
  {\bibfnamefont {J.}~\bibnamefont {Leuthold}}, \ and\ \bibinfo {author}
  {\bibfnamefont {J.}~\bibnamefont {Faist}},\ }\href
  {https://www.osapublishing.org/optica/fulltext.cfm?uri=optica-7-5-498&id=431641}
  {\bibfield  {journal} {\bibinfo  {journal} {Optica}\ }\textbf {\bibinfo
  {volume} {7}},\ \bibinfo {pages} {498} (\bibinfo {year} {2020})}\BibitemShut
  {NoStop}%
\end{thebibliography}

\begin{thebibliography}{7}%
\makeatletter
\providecommand \@ifxundefined [1]{%
 \@ifx{#1\undefined}
}%
\providecommand \@ifnum [1]{%
 \ifnum #1\expandafter \@firstoftwo
 \else \expandafter \@secondoftwo
 \fi
}%
\providecommand \@ifx [1]{%
 \ifx #1\expandafter \@firstoftwo
 \else \expandafter \@secondoftwo
 \fi
}%
\providecommand \natexlab [1]{#1}%
\providecommand \enquote  [1]{``#1''}%
\providecommand \bibnamefont  [1]{#1}%
\providecommand \bibfnamefont [1]{#1}%
\providecommand \citenamefont [1]{#1}%
\providecommand \href@noop [0]{\@secondoftwo}%
\providecommand \href [0]{\begingroup \@sanitize@url \@href}%
\providecommand \@href[1]{\@@startlink{#1}\@@href}%
\providecommand \@@href[1]{\endgroup#1\@@endlink}%
\providecommand \@sanitize@url [0]{\catcode `\\12\catcode `\$12\catcode
  `\&12\catcode `\#12\catcode `\^12\catcode `\_12\catcode `\%12\relax}%
\providecommand \@@startlink[1]{}%
\providecommand \@@endlink[0]{}%
\providecommand \url  [0]{\begingroup\@sanitize@url \@url }%
\providecommand \@url [1]{\endgroup\@href {#1}{\urlprefix }}%
\providecommand \urlprefix  [0]{URL }%
\providecommand \Eprint [0]{\href }%
\providecommand \doibase [0]{http://dx.doi.org/}%
\providecommand \selectlanguage [0]{\@gobble}%
\providecommand \bibinfo  [0]{\@secondoftwo}%
\providecommand \bibfield  [0]{\@secondoftwo}%
\providecommand \translation [1]{[#1]}%
\providecommand \BibitemOpen [0]{}%
\providecommand \bibitemStop [0]{}%
\providecommand \bibitemNoStop [0]{.\EOS\space}%
\providecommand \EOS [0]{\spacefactor3000\relax}%
\providecommand \BibitemShut  [1]{\csname bibitem#1\endcsname}%
\let\auto@bib@innerbib\@empty
\bibitem [{\citenamefont {Buhmann}(2012)}]{Buhmann2012a2}%
  \BibitemOpen
  \bibfield  {author} {\bibinfo {author} {\bibfnamefont {S.~Y.}\ \bibnamefont
  {Buhmann}},\ }\href {\doibase 10.1007/978-3-642-32484-0} {\emph {\bibinfo
  {title} {Dispersion forces I}}}\ (\bibinfo  {publisher} {Springer,
  Berlin/Heidelberg},\ \bibinfo {address} {Berlin},\ \bibinfo {year}
  {2012})\BibitemShut {NoStop}%
\bibitem [{\citenamefont {Leitenstorfer}\ \emph {et~al.}(1999)\citenamefont
  {Leitenstorfer}, \citenamefont {Hunsche}, \citenamefont {Shah}, \citenamefont
  {Nuss},\ and\ \citenamefont {Knox}}]{Leitenstorfer19992}%
  \BibitemOpen
  \bibfield  {author} {\bibinfo {author} {\bibfnamefont {A.}~\bibnamefont
  {Leitenstorfer}}, \bibinfo {author} {\bibfnamefont {S.}~\bibnamefont
  {Hunsche}}, \bibinfo {author} {\bibfnamefont {J.}~\bibnamefont {Shah}},
  \bibinfo {author} {\bibfnamefont {M.~C.}\ \bibnamefont {Nuss}}, \ and\
  \bibinfo {author} {\bibfnamefont {W.~H.}\ \bibnamefont {Knox}},\ }\href
  {\doibase 10.1063/1.123601} {\bibfield  {journal} {\bibinfo  {journal} {Appl.
  Phys. Lett.}\ }\textbf {\bibinfo {volume} {74}},\ \bibinfo {pages} {1516}
  (\bibinfo {year} {1999})}\BibitemShut {NoStop}%
\bibitem [{\citenamefont {Benea-Chelmus}\ \emph {et~al.}(2019)\citenamefont
  {Benea-Chelmus}, \citenamefont {Settembrini}, \citenamefont {Scalari},\ and\
  \citenamefont {Faist}}]{Benea-Chelmus20192}%
  \BibitemOpen
  \bibfield  {author} {\bibinfo {author} {\bibfnamefont {I.-C.}\ \bibnamefont
  {Benea-Chelmus}}, \bibinfo {author} {\bibfnamefont {F.~F.}\ \bibnamefont
  {Settembrini}}, \bibinfo {author} {\bibfnamefont {G.}~\bibnamefont
  {Scalari}}, \ and\ \bibinfo {author} {\bibfnamefont {J.}~\bibnamefont
  {Faist}},\ }\href {\doibase 10.1038/s41586-019-1083-9} {\bibfield  {journal}
  {\bibinfo  {journal} {Nature}\ }\textbf {\bibinfo {volume} {568}},\ \bibinfo
  {pages} {202} (\bibinfo {year} {2019})}\BibitemShut {NoStop}%
\bibitem [{\citenamefont {Moskalenko}\ \emph {et~al.}(2015)\citenamefont
  {Moskalenko}, \citenamefont {Riek}, \citenamefont {Seletskiy}, \citenamefont
  {Burkard},\ and\ \citenamefont {Leitenstorfer}}]{Moskalenko20152}%
  \BibitemOpen
  \bibfield  {author} {\bibinfo {author} {\bibfnamefont {A.~S.}\ \bibnamefont
  {Moskalenko}}, \bibinfo {author} {\bibfnamefont {C.}~\bibnamefont {Riek}},
  \bibinfo {author} {\bibfnamefont {D.~V.}\ \bibnamefont {Seletskiy}}, \bibinfo
  {author} {\bibfnamefont {G.}~\bibnamefont {Burkard}}, \ and\ \bibinfo
  {author} {\bibfnamefont {A.}~\bibnamefont {Leitenstorfer}},\ }\href {\doibase
  10.1103/PhysRevLett.115.263601} {\bibfield  {journal} {\bibinfo  {journal}
  {Phys. Rev. Lett.}\ }\textbf {\bibinfo {volume} {115}},\ \bibinfo {pages}
  {263601} (\bibinfo {year} {2015})}\BibitemShut {NoStop}%
\bibitem [{\citenamefont {Lindel}\ \emph {et~al.}(2021)\citenamefont {Lindel},
  \citenamefont {Bennett},\ and\ \citenamefont {Buhmann}}]{PRAlong2}%
  \BibitemOpen
  \bibfield  {author} {\bibinfo {author} {\bibfnamefont {F.}~\bibnamefont
  {Lindel}}, \bibinfo {author} {\bibfnamefont {R.}~\bibnamefont {Bennett}}, \
  and\ \bibinfo {author} {\bibfnamefont {S.~Y.}\ \bibnamefont {Buhmann}},\
  }\href {https://journals.aps.org/pra/abstract/10.1103/PhysRevA.103.033705}
  {\bibfield  {journal} {\bibinfo  {journal} {Phys. Rev. A}\ }\textbf {\bibinfo
  {volume} {103}},\ \bibinfo {pages} {033705} (\bibinfo {year}
  {2021})}\BibitemShut {NoStop}%
\bibitem [{\citenamefont {Lindel}\ \emph {et~al.}(2020)\citenamefont {Lindel},
  \citenamefont {Bennett},\ and\ \citenamefont {Buhmann}}]{lindel2020theory2}%
  \BibitemOpen
  \bibfield  {author} {\bibinfo {author} {\bibfnamefont {F.}~\bibnamefont
  {Lindel}}, \bibinfo {author} {\bibfnamefont {R.}~\bibnamefont {Bennett}}, \
  and\ \bibinfo {author} {\bibfnamefont {S.~Y.}\ \bibnamefont {Buhmann}},\
  }\href {https://journals.aps.org/pra/abstract/10.1103/PhysRevA.102.041701}
  {\bibfield  {journal} {\bibinfo  {journal} {Physical Review A}\ }\textbf
  {\bibinfo {volume} {102}},\ \bibinfo {pages} {041701} (\bibinfo {year}
  {2020})}\BibitemShut {NoStop}%
\bibitem [{\citenamefont {Boyd}(2003)}]{Boyd20032}%
  \BibitemOpen
  \bibfield  {author} {\bibinfo {author} {\bibfnamefont {R.~W.}\ \bibnamefont
  {Boyd}},\ }\href
  {https://books.google.de/books?id=3vHb7WGXmSQC&dq=nonlinear+optics+boyd&lr=&source=gbs_navlinks_s}
  {\emph {\bibinfo {title} {{Nonlinear optics}}}}\ (\bibinfo  {publisher}
  {Academic Press, San Diego},\ \bibinfo {year} {2003})\BibitemShut {NoStop}%
\end{thebibliography}
\end{document}